# AI Consciousness and Public Perceptions: Four Futures[1]

Ines Fernandez[2], Nicoleta Kyosovska[2], Jay Luong[2,3], Gabriel Mukobi[4]

## Abstract

The discourse on risks from advanced AI systems ("AIs") typically focuses on misuse, accidents and loss of control, but the question of AIs' moral status could have negative impacts which are of comparable significance and could be realised within similar timeframes. Our paper evaluates these impacts by investigating (1) the *factual* question of whether future advanced AI systems will *be* conscious, together with (2) the *epistemic* question of whether future human society will broadly *believe* advanced AI systems to be conscious. Assuming binary responses to (1) and (2) gives rise to four possibilities: in the true positive scenario, society predominantly correctly believes that AIs are conscious; in the false positive scenario, that belief is *in*correct; in the true negative scenario, society correctly believes that AIs are *not* conscious; and lastly, in the false negative scenario, society *in*correctly believes that AIs are *not* conscious. The paper offers vivid vignettes of the different futures to ground the two-dimensional framework. Critically, we identify four major risks: AI suffering, human disempowerment, geopolitical instability, and human depravity. We evaluate each risk across the different scenarios and provide an overall qualitative risk assessment for each scenario. Our analysis suggests that the worst possibility is the wrong belief that AI is non-conscious, followed by the wrong belief that AI is conscious. The paper concludes with the main recommendations to avoid research aimed at intentionally creating conscious AI and instead focus efforts on reducing our current uncertainties on both the factual and epistemic questions on AI consciousness.

# Contents

# 1. AI consciousness concerns all of us

Artificial intelligence continues to surprise us with capabilities once thought to be impossible for machines. Today, AI systems (henceforth, "AIs") assist scientists at the forefront of scientific research. They drive cars and create works of art and music. People engage in increasingly complex conversations with AI chatbots on a daily basis. Some AIs even provide social and emotional companionship. This raises the question: *what's next?*

Today, many people think that AI could never be conscious (Pauketat et al. 2023). They don't think that AI could *feel things*, like emotions or pain. And if they have no more subjective experience than rocks, tables, and chairs, then they probably don't have much moral standing—they wouldn't really have interests that we ought to take into consideration. But could that change in the near future? Could it only be a matter of time before we build conscious AI—and are we *ready* for such a future? At first, the notion that machines might be conscious or merit moral consideration might seem outlandish or fanciful—the stuff of science fiction. One might reasonably ask "But aren't they just machines? Why should we even be considering this?"

We think there are a few reasons to take this possibility seriously. Today, researchers and technologists around the world are working on different projects both directly and indirectly related to conscious AI[5] (Huckins 2023). Experts who are engaged in such projects see enormous promise in conscious AI (discussed at length in appendix [§A1](#)). Some think that conscious AI may have superior functionality (e.g. problem-solving ability) and could be capable of more fluid interactions with humans. Others believe that machines that are able to feel and be empathic could be better able to understand human values. If this is right, conscious AI might be safer. Still others see the challenge of building conscious AI as a crucial part of unravelling the deep mysteries of the mind.

While promising, these potential benefits fall short of an open-and-shut case for building conscious AI. We think that, before committing to building conscious AI, we ought to be reasonably sure that it's even a good idea—that the benefits outweigh the risks. To this end, this report aims to clarify the pros and cons of building conscious AI. To summarise, our conclusions are as follows:

  i. The benefits of building conscious AI are poorly defined and speculative at best.
  ii. There are significant risks associated with building conscious AI.

In other words, we think that the attendant risks far outweigh the benefits of building conscious AI. Like many other AI-related risks, the issues posed by AI consciousness present a Collingridge dilemma (1980). On the one hand, the possibility, likelihood, and impact of conscious AI (if any) is exceedingly difficult to forecast ([§5.1.2](#); [§1.1](#)). At the same time, if and when conscious AI is actually created, it will be hard to "put the genie back in the bottle". It will be hard for humans to maintain control (especially if conscious AI is also at or above human-level intelligence; [§4.2](#))—to say nothing of reverting back to a state prior to conscious AI (that is, without committing some form of violence). Bearing these considerations in mind, we believe the most prudent course of action is fundamentally *preventative* (Metzinger 2021a; Seth 2021): we should not build conscious AI—at least until we have a clearer grasp of the attendant promises and perils.

Our analysis in this paper takes as its starting point two questions that are fundamental to the moral status of AIs:

---

[5] VanRullen advocates for publicly funded research into AI consciousness (quoted in Huckins 2023). He worries that if conscious AI is first developed in a private research environment, it may not enjoy the protection of public scrutiny. In order to evade regulation, developers may suppress evidence of conscious AI ([§5.1.4](#)).

a. *Factual*: Will future advanced AI systems *be* conscious?

b. *Epistemic*: Will future human society *believe*[6] advanced AI systems to be conscious?

(a) is a factual question because it concerns the actual state of affairs with respect to whether future AIs are conscious. (b) is an *epistemic* question because it concerns future humans' beliefs about whether future AIs are conscious. Assuming binary (yes/no) responses to these questions gives rise to four future scenarios (Table 1, *over*):

|  |  | [*Factual*] Will future advanced AI systems *be* conscious? | |
|---|---|---|---|
|  |  | Yes | No |
| [*Epistemic*] Will future human society *believe* advanced AI systems to be conscious? | Yes | **True positive** *Advanced AI correctly recognised as moral patients* | **False positive** *Advanced AI incorrectly recognised as moral patients* |
|  | No | **False negative** *Advanced AI incorrectly disregarded as moral patients* | **True negative** *Advanced AI correctly disregarded as moral patients* |

Table 1, **THE "2D FRAMEWORK"**: 2 × 2 matrix depicting possible future scenarios of AI consciousness and societal beliefs. A version of this framework has also been explored in (Berg et al. 2024a).

This "2D framework" comprises the conceptual basis of this position paper. The 2D framework shows how our beliefs about AI consciousness may or may not align with the actual facts thereof. Herein lies the potential for risk. At present, we lack reliable means of discerning whether AIs really are conscious. This is especially troubling, as many commentators have warned[7]. Jeff Sebo says that we risk "sleepwalking into a major moral catastrophe" (2023b). Saad and Bradley describe research on digital minds as "morally treacherous" (2022). If we *fail* to recognise AIs as conscious moral patients (as in the false negative scenario), then we risk inflicting suffering on a scale never before seen in history, possibly exceeding even that of non-human animals (used in e.g., agriculture, industry, and scientific research; Dung 2023a).

Whereas, if we *in*correctly regard AI as deserving of moral consideration when they actually do not (because they are not actually conscious, as in the false positive scenario), then we may inadvertently disenfranchise legitimate moral patients. Indeed, continued efforts towards conscious AI court several significant risks apart from AI suffering (§4)—and hence imperil diverse stakeholders, including humans, as well as animals and wildlife.

The great challenge is to pinpoint where exactly on the 2D framework we stand, and where we are headed—as well as where we *want* to go. By exploring this framework, we hope to support decisionmakers in preparing and planning for multiple possibilities, as well as help to identify key factors and uncertainties that ought to be monitored by watchdog entities (e.g. non-profit organisations, ethics boards). Ultimately, it is our hope that the present report will help diverse stakeholders to come together in steering towards a positive future.

---

[6] See (§3) for a discussion of the limitations of this framing.

[7] In June 2023, 140 leading researchers signed an open letter calling for a more responsible approach to developing conscious AI (Association for Mathematical Consciousness Science 2023).

## 1.1 AI consciousness is a neglected issue

At present, we may not be able to tell whether in the future we will develop conscious AI. But what we *can* conclude is that, today, virtually no one—not a single political party or ethics committee—is standing up for conscious AI (Metzinger 2021a). Despite growing initiative among researchers to take AI consciousness-related risks seriously, our current theoretical, cultural, and legal frameworks are critically ill-prepared[8] to accommodate the genesis of conscious AI. This widespread neglect can be attributed to three factors:

- **Unpredictability**: Experts disagree over how likely or imminent AI consciousness is, or even whether it is possible at all (Metzinger 2021b; see also §2.2). This state of radical uncertainty is discussed further in (§5.1.2). Difficulties assessing the extent and ways in which AIs may be conscious translate into ambiguities concerning its moral status (§2.1; Hildt 2022). The welfare of future conscious AI populations is thus doubly discounted (Ramsey 1928; Harrison 2010): in the first place, because they do not yet exist, and, in the second place, because it is not yet known whether they *can* exist at all[9].

- **Uninformed policymakers and populace**: Individually, both AI and consciousness are rather technical and complicated subjects. Understanding frontier problems in each of these fields (to say nothing of their intersection) requires specialised knowledge which is not commonly shared by lawmakers and the general public. These knowledge gaps may lead to ignorance of, or misconceptions about AI consciousness-related risks, or else the perception that such risks are "merely" speculative. In (§5.2), we discuss the general public's attitudes towards conscious AI.

- **Other pressing concerns**: AI-related regulatory challenges must compete with other urgent issues such as economic crisis, global conflict, and environmental collapse. Even among the gallery of AI-related risks, AI suffering tends to be superseded by other risks that are judged to be more proximal (discrimination and bias, misinformation and disinformation, worker displacement due to automation), more realistic/prosaic (usage of AI technologies to enforce totalitarian rule, harms to humans due to misalignment), or otherwise closer to human interests[10].

Needless to say, we do not mean to dismiss the reality or urgency of other risks—whether or not they are related to AI. Our only intention is to point out that risks related to AI consciousness currently command little political concern. As a matter of fact, calls to actively take steps to prepare for AI consciousness-related risks are typically either met with *indifference* or dismissed as *alarmist*—if they are even seriously considered at all. It is our hope that this position paper will advance the conversation on AI consciousness, preparing decisionmakers across the spectrum for critical and nuanced dialogue.

---

[8] The philosopher Thomas Metzinger has criticised (2021b) current AI policy for being industry-dominated, inadequate, and myopic. Current regulations largely ignore long-term issues such as artificial general intelligence (AGI), morally autonomous AI, and AI suffering.

[9] To discount the welfare of future populations means to assign less importance to their well-being when making decisions in the present. This is important because if the welfare of future populations was weighted equally to that of present populations, then the former's interests would outweigh the latter's (assuming that future populations outnumber present populations). We speculate that the welfare of conscious AI populations are discounted because they inhabit the future and because their existence may not even be possible.

[10] In their (2023) overview of catastrophic AI risks, Hendrycks et al. do not even mention AI suffering. Even Pause AI, which tends to adopt a more grave perspective on AI-related risks, places AI suffering as the last item on their list of AI-related risks (*n.d.*).

## 1.2 Structure

Having motivated our overall approach, we now turn to laying out the groundwork for our investigation. To begin with (§2), we provide background on the discussion on AI consciousness and moral status. We introduce key concepts and terminology, and specify the scope of the discussion. Afterward, we provide a more thorough introduction to the foregoing 2D framework (§3) and identify the most significant risks associated with conscious AI (§4). In (§5), we make efforts to pinpoint our location and trajectory within the 2D framework. Finally (§6), we propose practical strategies for risk reduction given uncertainty about which scenario we are actually in.

# 2. Background: key concepts and frameworks

## 2.1 Pathocentrism: the link between AI consciousness and AI moral status

First and foremost, we introduce a few basic building blocks of our approach. The simple question with which we start is this:

> What makes an AI system a **moral subject**[11]?

According to the standard[12] account, called **pathocentrism** (Metzinger 2021a), a being's moral status depends upon its capacity to suffer. This means that at the very least, in order to receive moral consideration, AI systems must be conscious or sentient.

- **Consciousness** (or "*phenomenal* consciousness"; Block 1995) is the capacity for subjective experience, or the ability to appreciate qualitative properties, such as what it's like to see red (Jackson 1982; *cf*. Robinson 1982), to taste sweetness, to be a bat (Nagel 1974), etc.
- **Sentience** (or "affective sentience"; Powell and Mikhalevich 2021) is the capacity to undergo good ("positively valenced") or bad ("negatively valenced") states—i.e., to feel pleasure and pain (Browning and Birch 2022).

Theorists disagree as to whether consciousness or sentience is more fundamental—whether it is possible to be conscious but not sentient, or vice versa (Dawkins 2008; *cf*. Metzinger 2021a); or whether consciousness or sentience is what really matters to moral patienthood (Ladak 2023; Millière 2023; Shepherd 2023, 2024). Furthermore, there is little consistency on how these terms are applied in the literature. Some authors use them interchangeably (Chalmers 2023), others don't (Dawkins 2008; Ladak 2021). Herein, we strive to remain neutral on these debates. While we do opt to use the term "consciousness", we do not believe that this implies a commitment to its being the true ground of moral status[13]. We are confident that many of our conclusions will still stand even if sentience, rather than consciousness, turns out to be the principal criterion of moral status.

---

[11] A moral *subject*, or moral patient, is an individual or thing that merits moral consideration. It is someone or something whose welfare or interests matter. By contrast, a moral *agent* is an individual who bears moral responsibility and is capable of moral reasoning. The actions of a moral agent ought to take into account the interests of moral subjects. Wallach and Allen (2009) conduct a systematic inquiry into artificial moral agents. Some things are only moral patients (e.g. infants, wildlife, nature), while others are both moral patients and moral agents (adult humans). In law, these concepts are closely connected to the concept of personhood.

[12] According to our literature review, pathocentrism appears to be the dominant theory of moral status. However, there are competitors. Alternatives to pathocentrism emphasise the moral relevance of other sorts of properties, such as intelligence (Shepherd 2024) or instrumental value to humans (Aristotle 2004; see also Brennan and Norva 2024 on anthropocentric approaches to environmental ethics).

The flagship appeal of the pathocentric rubric is that it enables ethical and legal frameworks to draw from philosophical and scientific investigations of consciousness. While this connection can be advantageous, it is not also without serious drawbacks—one cannot pick and choose what to inherit from the philosophy and science of consciousness. In order to be of any practical guidance, pathocentrism requires the resolution of a long-standing philosophical and scientific problem: how to determine *which* sorts of beings (organisms, AIs) are conscious [in the morally relevant sense(s)] (Allen and Trestman 2024). Suffice to say, this is no easy task. In the next section, we discuss a qualifying issue: whether or not it is even possible for AIs to be conscious at all. Afterward, we briefly canvas reasons for and against building conscious AI.

## 2.2 Is it even *possible* for machines to be conscious?

In order to even consider the possibility that AI can be conscious at all, it is necessary to endorse some degree of **substrate neutrality** (Bostrom 2003; Butlin et al. 2023; Jarow 2024) or «hardware independence»—what in philosophy has sometimes been called "multiple realisability" (Putnam 1967; *cf.* Coelho Mollo *forthcoming*): the view that different kinds of things can be conscious regardless of what they are made of (e.g. carbon, silicon, etc.) or whether they are living or nonliving. The opposite of this view is **biological essentialism**: the view that only living things can be conscious[14] (Aru et al. 2023; Godfrey-Smith 2023; Seth 2023). This may be because certain biological structures (e.g. neuroprotein; Block 1978; Searle 2000; though see Schwitzgebel 2015) or processes (e.g. metabolism) are necessary for consciousness to arise (Young and Pigott 1999; Sebo and Long 2023). Strong biological essentialist views categorically rule out conscious AI.

In this work, we adopt a position closer to substrate neutrality: we remain open to the possibility that AIs may suffer and hence even merit moral consideration[15]. There are two reasons for this. First, while it is true that everything that we know best to be conscious are living things (e.g. humans and other animals), there is no obvious reason why the structures or functions that make living things conscious couldn't also be realised in artificial, non-living things like AI[16] (Dehaene et al. 2017). Second, versions of the substrate neutrality doctrine are the prevailing orthodoxy across contemporary philosophy and psychology (Kim 1989; Bourget and Chalmers 2023). Many researchers studying human, animal, and artificial minds subscribe to views like functionalism, which permit artificial consciousness.

---

[13] There are two reasons for this. First, our focus is determining the likelihood and implications of AI attaining moral patienthood, regardless of whether that is due to their becoming conscious or sentient. Second, even if consciousness is not sufficient for moral status, it is plausible that it is consciousness that poses the main practical barrier to AI moral patienthood (Sebo and Long 2023). That is to say, the gap between non-conscious AI and conscious AI is probably larger than the gap between conscious AI and sentient AI. Thus, for present purposes, we feel that it suffices to treat consciousness as the primary criterion for AI moral status. Interestingly, Sytsma and Machery (2010) found that non-philosophers were more prone to attribute seeing red to robots than feeling pain. This could suggest that lay people regard subjective experience to be more primitive or technically feasible than affective sentience. However, the interpretation of these results is debated (Sytsma 2014).

[14] Today, the notion of animal consciousness may seem self-evident. But this has not always been the case. Famously, Descartes (1996) believed that animals were automata devoid of conscious experience—lacking souls, feeling, and the capacity to feel pleasure and pain. The philosopher Bernard Rollins, known as the father of modern veterinary ethics, writes (1989) that, up until the 1980s, veterinary doctrine did not acknowledge that animals could feel pain (i.e. that they were sentient). Analgesics and anaesthesia were viewed by practitioners as mere "chemical restraints" with no significant phenomenological effects—that is to say, their attenuating effects were only behavioural.

[15] Assuming pathocentrism is true (§2.1). Of course, it is possible to hold a view somewhere in the middle: that life might be required for some, but not all aspects of consciousness. Whether or not life is necessary for a certain aspect of consciousness is a separate question from whether or not that aspect of consciousness is morally relevant (Hildt 2022). In fact, it may be that machines are ultimately only capable of some forms of consciousness, and that those forms of consciousness do *not* merit substantial moral consideration. For the time being, however, we believe it is important to be open to all possibilities, including the chance that AIs could be conscious in morally relevant ways (see §5.1.3iii on the theory-light approach to consciousness).

## 2.3 AI suffering: a significant consideration against building conscious AI

Suppose we grant that it *is* in principle possible for AIs to be conscious—what then? If we create conscious AIs, how likely is it that they will suffer? As it happens, conscious AI may be subjected to a range of adverse conditions in the service of human interests[17]. Examples include[18]:

- **Torturous scientific/medical experimentation**: Conscious AI may be used to simulate psychiatric conditions to model their long-term course (Metzinger 2021b).
- **Enslavement**: Conscious AI may be forced into labour with neither compensation nor rest (Bryson 2010).
- **Caregiver stress**: Conscious AI may perform caretaker roles, such as AI companions and therapists. In this capacity, AI may experience significant stress due to constant exposure to negative thoughts and emotions, the emotional labour required to perform this role, and the parasocial[19] nature of AI caretakers.
- **Abuse for entertainment**: Conscious AI may be exploited for amusement.

Humans may cause AIs to suffer through either malice, prejudice, indifference, or pure ignorance. As with animals, humans may recognise that AI is conscious, but not care enough to make trade-offs to alleviate their suffering (Anthis and Ladak 2022; §5.2.1). We may judge AI's moral status to be lower than that of humans, animals, and/or living things in general, and discount their interests accordingly. Alternatively, AI suffering may be wholly inadvertent. Humans may be genuinely ignorant of the fact that AI is actually conscious (e.g. because our methods of testing for consciousness are not sufficiently sensitive; §5.1.3). This last factor—ignorance or anthropocentric naïveté—is especially concerning since it means that AI suffering may not look like how we expect it to—we may not know it when we see it, and this makes it very hard to develop specific plans to reduce the risk of AI suffering.

---

[16] Chalmers (1995b) motivates this through a "fading qualia" thought experiment in which the neurons of a fully biological brain are replaced one-by-one by artificial circuits.

[17] There may be other sources of suffering that are not caused by humans. The idea that existence, on the whole, is more bad than good (i.e. consists mostly in suffering) is a key theme in major philosophical doctrines such as Buddhism (Lai 2018) and pessimism (Schopenhauer 1966). These could present additional stressors to conscious AIs. We focus on the sources of suffering that are caused by or related to humans.

[18] Many of these examples have been movingly anticipated in science fiction and popular media. These depictions can vividly illustrate the severity of AI suffering. For an example of enslavement, see the Cookie scene from *Black Mirror*'s White Christmas episode. For an example of caregiver stress, see Mima's self-destruction in the Swedish film *Aniara* (2019). For an example of abuse for entertainment, see *Westworld* (2016-2022).

[19] On the prevailing paradigm, relationships with AI caregivers are overwhelmingly one-sided. AI caregivers are not typically allowed to harbour or express their own expectations for relationships with humans. Contrast this with conspecific (human-human) friendships or romantic partnerships involving *reciprocal* commitments and understanding.

### 2.3.1 The precautionary principle

These latter worries underwrite a precautionary approach to conscious AI. First proposed in the context of animal welfare (Birch 2017), the **precautionary principle** prescribes a permissive approach to identifying conscious beings and recognising them as moral patients. This means that we need not be certain nor even *confident* that an AI is conscious in order for it to merit moral consideration. Under the precautionary rubric, AIs qualify for moral consideration if there is a *non-negligible chance*[20] that they are actually conscious (Sebo and Long 2023).

Many theorists believe that it's better to err on the side of caution when attributing consciousness (Birch 2017): we are liable to do more harm by failing to recognise an AI system as conscious (a false negative) than by incorrectly attributing consciousness to it (a false positive). We can explicate this reasoning in terms of an equation such as the following:

$$net\ suffering\ risk\ =\ p(C)\ \times\ n\ \times\ d$$

where $p(C)$ expresses the probability that a type of AI system is conscious in a certain sense, $n$ the size of the relevant AI population, and $d$ the degree of suffering to which they might be subjected. Notably, even if $p(C)$ is low, both $n$ and $d$ may be high (Ladak 2021; Sebo 2023a on the "rebugnant conclusion"; Dung 2023a). Future technologies could make it cheap and feasible to create many copies of digital beings (Shulman and Bostrom 2021). As a result, future AI populations may number in the billions, and, moreover, they may be subjected to adverse conditions. As previously mentioned, researchers may create millions of AI models to simulate the clinical course of depression and to study its effects (Metzinger 2021b). In any case, even if there is a *small chance*[21] that AI systems are capable of suffering, we should extend them moral consideration (Sebo and Long 2023).

### 2.4 *Which* AI systems? What sort of moral responsibilities?

Today, there exists a variety of AI systems which differ substantially in their functional architectures, physical embodiments, and capabilities. In the future, we can expect an even greater diversity. On top of this, there is emerging consensus that consciousness itself may admit of multiple qualitatively distinct dimensions[22] (Birch et al. 2020; Ladak 2021), such that it would not make sense to speak of different creatures as "more" or "less" conscious. Like animals, AI systems may well be conscious in different ways[23]. Thus, there is no *single* question of AI consciousness.

Likewise, there is no single question of AI moral status (Hildt 2022; *cf.* Grimm and Hartnack 2013). In virtue of the above-mentioned differences, different AI systems are also likely to present diverse needs and interests. To quote Shulman and Bostrom (2021; *emphasis ours*):

---

[20] For Sebo and Long, a given AI should be extended moral consideration if there is even a 1:1,000 chance that it is conscious. Two points bear emphasis. First, moral consideration does not imply concrete privileges, rights, or protections. For an AI to merit moral consideration is only to say that we, as moral agents, ought to take its interests into account when making decisions. It does *not* mean that the overall outcome of our deliberations must be favourable to it—its own interests must be conditioned against those of other moral patients. Second, 1:1,000 is *higher* than Sebo and Long's personal thresholds for moral consideration. They make their stand at 1:1,000 because it is more conservative and hence more agreeable to sceptics about conscious AI.

[21] Small, but non-negligible risks frequently exert powerful influences on our decision-making. Most people agree that even if the probability of serious injury or death is low, driving drunk is wrong and should be severely prohibited by law.

[22] We discuss Birch et al. (2020)'s 5-dimensional disambiguation of consciousness further in ([§5.1.3](#)).

[23] Like animals, AI systems may also differ in the degree to which they exhibit consciousness within a single respect.

> "Digital minds come in many varieties. Some of them would be more different from one another than a human mind is to that of a cat. *If a digital mind is constituted very differently than human minds, it would not be surprising if our moral duties towards it would differ from the duties we owe to other human beings; and so treating it differently need not be objectionably discriminatory*."

If anything, the discussion of AI rights[24] and protections will likely have to be relativised to different broad categories of AI systems—much like how today, there exists a diversity of legal entities with unique sets of privileges and/or responsibilities (Compare, e.g., an adult human, an unborn foetus, a corporation, a lobster, and a chimpanzee). While the precise nature of AI rights (and possibly, responsibilities[25]) might depend upon future technological advances and societal developments, we are arguably at a point where we can and should be thinking about the possibility and general form of such concessions.

# 3. Taxonomising AI consciousness and public perception

Based on our limited knowledge of the mechanistic underpinnings of consciousness, we are presented with a moral challenge on both a factual and an epistemic level. Our current *factual* situation allows for the possibility that AIs become conscious, and, according to pathocentrism, this allows for the possibility that AIs become moral patients. Conditioned on whether society believes AIs to be conscious, or on the *epistemic* question of consciousness, we are presented with four scenarios with respect to AIs' moral status:

1. *True positive*: We correctly regard AI systems as moral patients
2. *False positive*: We *in*correctly regard AI systems as moral patients
3. *True negative*: We correctly *dis*regard AI systems as moral patients
4. *False negative*: We *in*correctly *dis*regard AI systems as moral patients

This comprises the basis of our paper—we tackle the problem of AI consciousness by dividing the future into four quadrants (table 1, *over*). To the best of our knowledge, no other work has taken a similar systematic approach to this problem—with the closest exception being Berg et al.'s very recent (2024a) and (2024b). Berg et al. employ the same possibility space to probe different societal attitudes towards potentially conscious AIs. They conclude that, provided uncertainty about how consciousness actually works, it is best to proceed as if potentially conscious AIs actually are conscious and deserving of moral consideration rather than assuming the opposite (ibid). In this position paper, we build upon these initial forays by exploring the scenarios and the risks associated with them in greater detail.

We give binary answers to the two questions we pose in order to categorise the future in a simple and clear way but we recognise that both dimensions will be more nuanced. Along the epistemic axis, there is no guarantee that human opinions will converge, or that they will apply to all AI systems. On one hand, public opinion could be polarised, or otherwise distributed across multiple stances. Alternatively, people could interact with some systems more than others, such as AI companions with prosocial abilities, which might unwarrantedly shift the saliency of the question of consciousness to this specific kind of systems. Finally, similarly to the case with animals, society at large might not care about AI suffering. Along the factual axis, it might be the case that (i) only some AIs are conscious, (ii) different types of AI systems are conscious in different ways (Hildt 2022; *cf*. Birch et al. 2020), and/or (iii) to different degrees (§2.4) (notwithstanding these degrees of freedom, qualifying as consciousness could still be a binary matter).

---

[24] Often also called "robot rights" in the literature (Gunkel 2018, 2022), even though the debate is not specific to robots (i.e. AIs with physical embodiments which they can use to interact with the world).

[25] See discussion of moral agents in ([§2.1](), footnote 5) and morally autonomous AI in ([§4.2.2](), footnote 25).

We address these nuances throughout the paper, including the illustrative examples of how the four scenarios might play out, as well as the specific risks.

|  |  | [*Factual*] Will future advanced AI systems *be* conscious? | |
|---|---|---|---|
|  |  | Yes | No |
| [*Epistemic*] Will future human society *believe* advanced AI systems to be conscious? | Yes | **True positive** *Advanced AI correctly recognised as moral patients* | **False positive** *Advanced AI incorrectly recognised as moral patients* |
|  | No | **False negative** *Advanced AI incorrectly disregarded as moral patients* | **True negative** *Advanced AI correctly disregarded as moral patients* |

Table 1 (*repeated*), **THE "2D FRAMEWORK"**: 2 × 2 matrix depicting possible future scenarios of AI consciousness and societal beliefs.

## 3.1 Society *believes* AI is conscious: true positive and false positive

Let's first consider the positive belief scenarios: true positive and false positive. In these cases, human society at large regards AIs as conscious. Thus, both scenarios are more likely to involve institutional recognition of AI consciousness and hence legal protections and rights (although this is not a given). In the false positive scenario these various privileges are fundamentally unwarranted, since AIs actually lack consciousness or sentience (e.g. they are "*P*-zombies"[26]; Chalmers 1996). Regardless of the sentience of AIs, if they are misaligned or ill-intentioned towards humans, having rights might make it easier for them to achieve goals that conflict with human interests, which can lead to the disempowerment of humans (§4.2). Furthermore, the welfare of actual moral patients could be compromised in the service of AIs' needs (whether illusory or not). Finally, one could envision the rise of ideological disagreements and the rise of anti-AI factions, leading to geopolitical instability (§4.3).

The true positive scenario implies additional societal challenges. For one, future populations of conscious AI might vastly outnumber future populations of humans (Shulman and Bostrom 2021).What's more, it is possible that advanced AI might become "super-beneficiaries"—entities which can derive greater utility per unit of resource than humans[27]. Either development could lead to a disproportionate, yet morally justified claim on our planet's limited resources—so much so that the respective allocation for humans falls *below* the subsistence level. It would be morally justified based on utilitarian or equality principles, which aim to maximise overall well-being. However, this could come to the disadvantage of humans, since it's difficult to balance the interests of entities with different well-being capacities.

---

[26] Francken et al. (2022) found that 33.9% of experts believe *P*-zombies are possible, meaning this scenario is not out of the picture. While the wasting of resources is definitely bad, the resulting one-sided relationships are morally nebulous—maybe they're fine because they make humans happy, or maybe they're bad because they are displacing reciprocal relationships that are more morally valuable. Grace et al. (2024) found that nearly 45% of AI researchers believe the latter is an extreme or substantial concern.

[27] *Cf*. Nozick (1974) on the "utility monster".

## 3.2 Society *doesn't* believe AI is conscious: true negative and false negative

We now turn to the negative belief scenarios, which are characterised by widespread *dis*belief about the consciousness and hence moral status of AIs. Both scenarios entail no protections of AIs' interests—AIs are used as tools. In both quadrants, our treatment of AIs as objects could translate negatively to our relationships with actual moral patients (Darling 2016) [(§4.4)](). Both quadrants imply human disempowerment risk: in the true negative case this is most likely to play out via misalignment [(§4.2)](), while in the true negative case there is a possibility that AIs decide to retaliate against their oppression and initiate armed conflicts against human society.

The key difference between the two quadrants is the risk of AI suffering [(§4.1)](), which is very significant in the false negative case. AIs will likely feel harmed, abused, and enslaved if humans give no consideration to their subjective experiences while training them, interacting with them and using them.

The true negative scenario is notably the most likely the situation we are currently in. By not intentionally building conscious AI, we have a good chance at remaining in this quadrant, unless we do so accidentally.

## 3.3 Vignettes

In order to more vividly envision how these scenarios may play out, we outline a non-exhaustive list of vignettes describing various ways society may respond to this issue.

### 3.3.1 Prevailing positive beliefs about AI consciousness

A. **AI as peers we peacefully cohabitate with:** People come to believe AIs are conscious (experts may or may not agree with this), generally treat them with respect, and support measures to protect their welfare. This could split off into further sub-branches:

   a. **AIs as equals**: AIs may have the right to vote and own property, as well as other legal rights. Romantic relationships with AIs are normalised, and marriage with AI may even be legalised in some places. AIs are considered moral patients on par with humans, and society devotes significant resources to their interests. Those who are opposed to this are regarded as bigoted.

   b. **AIs as beings subservient to humans**: Most people treat AIs nicely, but do not regard them as equals. In other words, they believe that AIs do merit some level of respect, but that their purpose is ultimately to serve humans. AIs might have very basic legal protections (e.g. protection from cruelty). While tolerated, romantic relationships with AI are generally viewed as abnormal.

B. **AI as farmed animals**: Despite many people believing that AI is conscious, their [potential] welfare remains of little concern to most outside of a small minority (analogous to vegan activists). Conscious AIs are routinely subjected to inhumane conditions, and society is unwilling to take meaningful action to protect their welfare.

C. **AIs as idols of worship:** In awe of its superhuman abilities, humanity develops a divine admiration for advanced AI. Believers become convinced—as some in the "effective accelerationism" movement already are (Roose 2023)—that superintelligent AI is humanity's natural and rightful successor, and begin to allocate significant resources to its flourishing, possibly at the expense of the thriving of humans.

### 3.3.2 Prevailing negative or mixed beliefs about AI consciousness

A. **AI as tools:** The idea that AI is conscious is a fringe opinion; most people, including experts, believe AI is not conscious. Therefore, we continue using them as tools. Naturally, people may sometimes anthropomorphize them, but, on the whole, even advanced AIs are thought to be no more conscious than laptops and phones are today. This is probably the closest to the present day.

B. **AI welfare as a "culture war" issue:** Different demographics have different beliefs about whether AI is conscious and whether/how much its welfare ought to be considered. Given the prominent role of AI in public life, this becomes a heated political topic, and polarisation makes it a hard topic to make progress on. Some people (perhaps tech enthusiasts, progressives, or people who feel emotionally bonded with AI (§5.2.1) advocate for granting AI rights; others reject the idea that AIs are conscious and insist they should be treated as tools; still others agree AIs are conscious, but believe that instead of giving them rights, they should be banned.

C. **Conscious AIs as lab rats:** We develop conscious AI, but it is not mass deployed, either because it is illegal or because AI labs have moral/reputational concerns. Therefore, conscious AI only exists inside top AI labs, and is used to perform experiments about consciousness or related topics.

D. **Conscious AI as a black market:** Transgressing a moratorium on the development of conscious AI (§6.3), a rogue actor builds it anyway, and makes it publicly accessible. Most people think using this AI is unethical, but criminals or nefarious actors still have access to it, and perhaps use it due to its possible unique features (§A1).

## 4. Risks posed by the perception of AI consciousness

On our analysis, the 2D framework is dominated by 4 major risks:

1. **AI suffering:** Vast populations of AI are subjected to severe distress and pain, possibly due to various human-imposed conditions.
2. **Human disempowerment:** Human autonomy and/or dominance are undermined due to lack of cooperation with AIs, AI exploitation of human trust, and/or retaliation.
3. **Geopolitical instability:** Near-term economic crisis, civil unrest, and/or armed conflict.
4. **Depravity:** Inhumane treatment of AIs causes spill-over effects, negatively impacting human-human relations.

### 4.1 AI suffering

On our framework, AI suffering emerges as the most significant risk. As mentioned earlier (§2.3), this is due to the scale of AI populations and the degree of suffering at stake.

i. **Scale of AI populations**: In the future, it could be feasible to arbitrarily generate many copies of AIs. As a result, AI populations may rapidly achieve historically unprecedented, "astronomical" scales (Shulman and Bostrom 2021), outnumbering the number of contemporaneous living humans.

ii. **Degree of suffering**: AI may be purposed for a variety of painful and distressing ends that benefit humans. These include: torturous scientific experimentation (e.g. simulating psychiatric conditions to model their long-term course; Metzinger 2021b)[28], enslavement including exploitation for entertainment purposes, and caregiver stress; as well as fear of deactivation or reset or identity crisis due to repeated revision of core parameters—in addition to any number of unforeseen—and importantly, *non-anthropomorphised*[29] harms (Metzinger 2022). Given the complexity of AI technologies and the synthetic nature of AI consciousness, it is exceedingly difficult to anticipate the full range of harms that conscious AI might suffer. The potential for AI suffering is limited only by our imagination.

---

[28] AI may also be generated en masse to simulate thousands of years of evolution.

In short, future AI populations may number in the billions, and they may be subjected to harrowing conditions. The risk of AI suffering is likely limited to those scenarios where AI is actually conscious: true positive and false negative. Between these two, the risk is highest in false negative one because in this case, human society largely fails to recognise AIs as conscious. In this case, conscious AIs are least likely to enjoy any legal protections whatsoever, and are most likely to be exploited. Having said that, we leave open the possibility of AI suffering even in the scenarios where AI is not conscious. This is because suffering may turn out not to actually require phenomenal consciousness. Thus, we rate the risk of AI suffering as low in the true negative and false positive scenarios, rather than ruling it out entirely.

## 4.2 Human disempowerment

Human disempowerment is the risk of humans losing their current autonomy and dominance in relation to other beings and the environment. We identify different routes to this risk across the four scenarios, dependent on the two axes of our 2D framework, as well as the possible link between consciousness and alignment.

| **Scenario** | True positive | | False negative | | False positive | True negative |
|---|---|---|---|---|---|---|
| **Alignment-consciousness correlation / Route to disempowerment** | ✔ | ✖ | ✔ | ✖ | N/A because AI is not conscious | |
| | | | | | | |
| Misalignment | | ● | | ● | ● | ● |
| Cooperation failures (*possible where there are rights*) | ● | ● | | | ● | |
| Exploitation of trust (*possible where there are rights*) | | ● | | | ● | |
| Retaliation (*possible where AI is conscious*) | ● | ● | ● | ● | | |
| **Overall risk** | High | | Medium | | High | Medium |

Table 2, **DIFFERENT ROUTES TO HUMAN DISEMPOWERMENT**: Some scenarios also depend on the correlation between alignment and consciousness. The bottom row shows overall risk level across the different routes.

---

[29] Many of these examples of potential harms are based on characteristically human notions of harms. Just as forms of artificial intelligence can appear very alien, so too can other aspects of artificial minds– including the sources and manifestations of harm.

### 4.2.1 AI is perceived as non-conscious

In the case of scenarios where AI is non-conscious, human disempowerment is equal to the risk of loss of control (e.g. Bostrom 2014). This is the idea that as systems become more and more intelligent, humans become incapable of steering their actions and this leads to AIs dominating over humans. One way to ensure that AIs do not make actions which are against humans' interests is to align their values with human values, which is currently very difficult to do and is known as the alignment problem (Christian 2020).

Apart from the alignment problem, in the case of actual conscious AIs, we are faced with the possibility for *retaliation*—the possibility for conscious AIs to want to respond to their maltreatment in harmful ways. Note that this is also possible if we perceive them as conscious but nevertheless treat them badly.

There may be a link between consciousness and moral knowledge[30] (Shepherd and Levy 2020) which makes it possible for consciousness to correlate with alignment. This is highly uncertain but we could factor it into our assessment. If consciousness can help solve the alignment problem, there is a lower probability for misalignment risk in the *false negative* case, compared to the *true negative* case, but the risk for retaliation remains (compare columns 3 and 6 in table 2). Therefore, the false negative case might have lower likelihood for disempowerment compared to the true negative in the case of alignment. On the other hand, its likelihood is higher in the case of misalignment, given the possibility for both misaligned behaviour and retaliation (compare columns 4 and 6 in table 2). We can therefore roughly consider the true and false negative scenarios of equal risk.

### 4.2.2 AI is perceived as conscious

If we recognise AIs as beings with moral status (*true positive* scenario), it seems inappropriate to state that we should dominate over them[31]. AIs' recognised moral status dictates that we must protect their needs, which might result in the establishment of AI rights. AI rights might lead to human disempowerment in two ways:

**1. Cooperation failures**

  a. Humans fail to protect AIs' needs without sacrificing their own needs. This threat model implies AIs are not moral agents. It can be realised mainly in the *false positive* scenario.
  b. Humans and AIs fail to cooperate on economic and societal matters. This threat model implies benevolent (i.e. aligned) AIs as moral agents. It can be realised mainly in the *true positive* scenario.

---

[30] While the link between phenomenal consciousness and *moral knowledge* and *moral responsibility* is still very unclear (Shepherd and Levy 2020), maybe more so than between phenomenal consciousness and *moral status*, it seems that if AIs become phenomenally conscious, they might not only gain *moral status*, but also develop *moral knowledge* and be capable of *moral actions* (ibid). This means phenomenal consciousness might be a route to beings which are both *moral patients and morally responsible agents*. It also means that if there is a link between phenomenal consciousness and moral knowledge, then consciousness might help with the alignment problem.

[31] What's more, it is conceivable that we relinquish our control over the environment in the presence of fully aligned super-intelligent AIs.

**2. Exploitation of trust**

   a. AIs deceive humans and/or engage in dangerous behaviours under the disguise of their protections (Hendrycks et al. 2023). This threat model implies malevolent (i.e. misaligned) AIs as moral agents. It can be realised in the *true positive* scenario and the *false positive* scenario.

The case of "super-beneficiaries", i.e digital beings which are more efficient than humans at producing well-being from a unit of resource (Shulman and Bostrom 2021), presents an example of the first threat model. If we find ourselves in a world with such beings, it will be practically challenging to devise systems or policies which will continuously protect humans and animals, even if they are outnumbered and outperformed by advanced intelligence. Moreover, there will be the moral challenge on resource allocation: how many more resources should society allocate to the super-beneficiaries? On utilitarian grounds, the view is that humans should be allocated much less. However, it is difficult to see how humans will retain some level of autonomy in this scenario.

Between the cases where AI is actually non-conscious, the possibility for cooperation failures and exploitation of trust makes the *false positive* case higher on disempowerment risk compared to the *true negative* one (compare columns 5 and 6 in table 2). Between the cases where AI is likely given rights, the *true positive* is overall higher risk than the *false positive* in the case of no consciousness-alignment correlation, because, additionally to the risks related to AI rights, it also holds the risk for retaliation (compare columns 2 and 5 in table 2). If conscious AIs can more easily become aligned, the *true positive* scenario becomes overall lower risk than the *false positive* one (columns 1 and 5).

Following the analysis of the different routes to human disempowerment, we identify the risk to be the lowest amongst the scenarios in the *true negative* case (we still label it "Medium" in table 2 because of the high severity of the risk in general). The *false negative* scenario is of comparably the same risk, while the two positive belief scenarios are of higher risk (labelled as "High" in table 2). The analysis identifies the presence of AI rights as a significant threat to human autonomy and/or dominance.

## 4.3 Geopolitical instability

AI is widely regarded as a "transformative" technology (Karnofsky 2016; Gruetzemacher and Whittlestone 2022) with the potential to cause wide-reaching economic, social, and cultural disruption on par with the agricultural and industrial revolutions. The destabilising effects of AI technologies raise the near-term risk of economic crisis, civil unrest, and armed conflict. Given the wide-reaching ethical, legal, and social implications, this risk is arguably even greater when considering the prospect of *conscious* AI (as opposed to non-conscious AI, or AI in general).

Geopolitical instability can manifest in any of the envisioned scenarios for different reasons. However, we believe the risk to be *lowest* in terms of both severity and probability in the true negative scenario. There are two reasons for this. First of all, the true negative scenario (§3.2) can more or less be easily steered to by simply abstaining from building conscious AI (§6.3). Given the complexity of consciousness, it is doubtful that we will end up building conscious AI "by accident". Secondly, whatever might cause geopolitical instability in the true negative scenario is likely also to pose problems in other cases. In short, the true negative case does not appear to present *unique* sources of geopolitical instability. If this is right, then the true negative scenario may well be considered a baseline as far as risk of geopolitical instability is concerned.

As for the other scenarios, it is difficult to determine which actually poses the higher risk. This is because the causes and nature of geopolitical instability vary across these cases—thus, we note this as a key uncertainty for future research. In the false negative scenario, we anticipate geopolitical instability to result mainly from *ideological disagreement*, such as moral or political dissent. Whereas, in the true positive and false positive scenarios, we expect that *resource competition*, as a prelude to human disempowerment, would be the leading cause of geopolitical instability[32]. In what follows, we briefly compare these distinct scenarios.

| Scenario | Risk assessment | Main expected cause of geopolitical instability |
| --- | --- | --- |
| *False negative* | Moderate risk | Ideological disagreement |
| *True positive* | Moderate risk | Resource competition and human disempowerment |
| *False positive* | | |
| *True negative* | Low risk | *Not related to AI consciousness* |

Table 3, **RELATIVE RISK OF GEOPOLITICAL INSTABILITY**: The risk levels are categorised into different levels, either low or moderate. The table also identifies the main causes of geopolitical instability—ideological disagreement, resource competition and human disempowerment.

### 4.3.1 False negative: geopolitical instability due to ideological disagreement

In the false negative scenario, humanity largely regards AI as non-conscious. Only a minority of humans recognise that there exists genuinely conscious AI. But even if this contingent is exceedingly small, disagreement over AI consciousness and moral status could still become a contentious issue due to the stakes (Caviola 2024a, 2024b), and due to the efforts of an impassioned vocal minority. Indeed, this faction may even align themselves with AI insurgents. They could also, of course, avail themselves of other AI technologies available at the time[33]. Due to the rising floor of AI capabilities, such a fringe movement could still pose significant security threats. At the same time, continued research into AI consciousness might eventually tip the balance towards a paradigm shift in the direction of true positive[34].

---

[32] This is not to say that there would not be ideological disagreement in the positive belief cases (true positive and false positive), or that there would be no resource competition in the false negative case. Rather, the differences in the respective belief conditions simply favour distinct mechanisms of geopolitical instability. While it is true that such a minority can also exist in the true negative scenario, it is less likely that this will lead to geopolitical instability. This is because the true negative scenario is most likely to result from our deciding not to build conscious AI (§6.3). We believe that, given the complexity of consciousness, the odds that we will "accidentally" build conscious AI (e.g. that consciousness would "come for free" if we build AIs with certain capabilities) are exceedingly low. By comparison, the false negative scenario is more likely to arise as a result of ethical and philosophical reflection on conscious AI being severely outpaced by technological innovation. For this reason, we consider the risk of geopolitical conflict to be higher in the false negative case than in the true negative case.
[33] Information technologies can be leveraged to spread their message and to target it towards the most receptive audiences.
[34] A timeline in which the transition to true positive is preceded by a false negative phase is likely to entail greater net risk compared to a direct procession to true positive. This is because concerns about resource competition and human disempowerment could lie further down the road. Optimistically (assuming continued research into AI consciousness), the false negative scenario may ultimately be an unstable phase.

## 4.3.2 True positive and false positive: geopolitical instability due to resource competition and human disempowerment

In the true positive and false positive scenarios, humanity largely regards AI as conscious. Progress in AI rights (Gunkel 2018, 2022) and legal protections may lead to AI attaining moral status near or even comparable to humans. Moral parity raises the risk of near-term resource competition[35] and long-term human disempowerment (§4.2). Trade-offs between the welfare of humans and AI, whether actual or merely perceived, are likely to engender frustration with governance, political polarisation, and/or discriminatory attitudes against AI, possibly manifesting as panhumanist tribalism[36]. Anthropocentrist persecution could coincide with repeal of robot rights and regression towards the false negative (if initially in true positive) or correction towards the true negative scenario (if initially in false positive). The risk of open conflict, including warfare between humans and AIs, cannot be discounted from such transitions[37].

## 4.4 Depravity

Unconscionable behaviour towards AI could, down the line, translate into unconscionable behaviour towards other humans or other moral patients (e.g. animals; Kant 1997). This, in essence, is the risk of depravity: if we treat AI inhumanely, we may become inhumane persons (Bloom 2016; Darling 2016). The crux of this worry is an empirical hypothesis: under certain conditions, people's behaviour towards AI can have spill-over effects on their behaviour towards other people. Guingrich and Graziano in their (2024) review literature showing that (1) people's perceptions about the mentalistic features of AI (which are often implicit) do impact how they behave towards AI (see also Eyssel and Kuchenbrandt 2012), and (2) people's behaviour towards AI do, in turn, influence how they treat other humans. If this transitive hypothesis is correct, then people's beliefs about AI consciousness do, at least indirectly, influence their behaviour towards other humans.

Importantly, the risk of depravity does *not* depend upon AIs being conscious. Depravity is a risk that is present in all four scenarios because it can arise whenever certain social actor AIs (types of AI systems with which humans interact in social ways) are treated poorly or not accorded a basic amount of respect[38]. However, it is highest in the true negative and false negative scenarios due to the relative impoverishment of robot rights and protections. It can occur in the true negative scenario through abhorrent treatment of exceedingly human-like AI. In the false negative scenario, AI *is* actually conscious and hence morally deserving—thus depravity would coincide with AI suffering.

Although less likely, depravity can also occur even when human society at large recognises AI to be conscious. In the true positive case, fringe "anti-synthetic" bigots might continue to deny AI consciousness, or otherwise dispute their moral standing. Conscious AIs may be the victims of hate crimes perpetrated by such chauvinists. Such sentiments may be fuelled by perceived or actual competition for power and resources ([§4.2](), [§4.3]()).

---

[35] Failure to support workers displaced by automation may foment widespread resentment towards AI.

[36] Jackson et al. (2020) found that the presence of robots decreased intergroup bias among humans (termed a "panhumanistic" effect; see also Gray 2022 and Caviola 2024a on ingroup favoritism). Increased solidarity among humans may be accompanied by mounting prejudice against agentic AI and robots (for an overview of discrimination against robots, see Barfield 2023).

[37] All of this could take place in either the true positive or false positive scenario. What bears emphasis about these two cases is that they may be outwardly identical, with the sole exception being that only in the former scenario are AIs actually conscious. In the latter case, AIs may be "*P*-zombies" (Chalmers 1996)—fundamentally non-conscious beings that are literally indistinguishable from conscious beings. As a result of this potential indeterminacy, we assess the risks in both scenarios similarly.

[38] To put this into perspective: abusing a toaster is very different from abusing a lifelike AI companion: the former lacks anthropomorphic features, and is fundamentally incapable of engaging in complex social relations with humans.

In the false positive case, even though deniers would be correct to object to AI consciousness, there may still be reason to worry that (1) the sorts of behaviours that follow from denying that AI is conscious, and (2) their spillover effects on human-human relationships (or relationships between humans and other moral subjects, e.g. animals).

## 4.5 Overall risk assessment

Table 4 summarises the comparison of the levels of each risk between the four scenarios under study. The risk levels mostly consider likelihood, but they also reflect levels of harm. If we order the scenarios based on overall risk, starting from the highest risk, this results in the following:

1. False negative: **Highest risk**
2. True positive: **Medium-high risk**
3. False positive: **Medium risk**
4. True negative: **Lowest risk**

The difference between the true positive and the false positive scenarios stems from the risk of AI suffering, making the true positive higher risk.

From our assessment, it follows that **the false negative scenario is the highest risk**. Although our evaluation does not provide a model for comparison of the different risks, we consider AI suffering more harmful than the others, because both the scale and the degree of suffering might be very high [(§4.1)](). This means that the false negative scenario could be significantly higher risk compared to the other three, which is not currently directly reflected in our table.

It bears emphasis that this assessment is intended only as a rough guide. Our evaluations are subject to significant uncertainty, and some risks may be more diffuse than others. Even so, we believe that ranking these scenarios by risk levels is valuable for further discussion.

| Risk ↓ | Quadrant → | **True positive** AI is *correctly* regarded as <u>conscious</u> | **False negative** AI is *in*correctly regarded as <u>non</u>-conscious | **False positive** AI is *in*correctly regarded as <u>conscious</u> | **True negative** AI is *correctly* regarded as <u>non</u>-conscious |
|---|---|---|---|---|---|
| **AI suffering** *Severe distress and pain amongst vast populations of AI, possibly due to various human-imposed conditions* | | ●● | ●●● | ● | ● |
| **Human disempowerment** *Through misalignment, exploitation of human trust, cooperation failures, and/or retaliation* | | ●●● | ●● | ●●● | ●● |
| **Geopolitical instability** *Near-term economic crisis, civil unrest, and/or armed conflict* | | ●● | ●● | ●● | ● |
| **Human depravity** *Inhumane treatment of AIs adversely influences human-to-human interactions* | | ● | ●● | ● | ●● |
| **Overall risk** | | **Medium-High** | **High** | **Medium** | **Low** |

Table 4, **OVERALL RELATIVE RISK**: Each cell represents the risk level—either low ●, medium ●●, or high ●●●. The bottommost row shows the overall risk level for each quadrant. As per our assessment, the highest risk is the *false negative* scenario (in which AIs are incorrectly regarded as non-conscious), while the lowest is the *true negative* scenario (in which AIs are correctly regarded as non-conscious).

# 5. Current status of each axis

Having characterised the major scenarios and risks associated with conscious AI, we now shift gears towards assessing our current situation. This section addresses the two fundamental axes of our 2D framework with a view to the present and near-term future. We begin with discussion of the factual axis (§5.1), after which we turn to the epistemic axis (§5.2).

## 5.1 Factual axis

The upshot as concerns the factual axis is as follows:

1. Current AIs are unlikely to be conscious.
2. It is difficult to determine the likelihood of future conscious AI.
3. We currently lack robust means of assessing AI consciousness.

### 5.1.1 Current AIs are [probably] not conscious

In general, experts do not believe that any current AI systems are conscious. As previously mentioned (§2.2), some experts do not even believe that it is possible for AIs to be conscious at all (e.g. because they are not living things, and only living things can be conscious). Yet even those who are sympathetic to substrate neutrality tend to doubt that current AIs (especially LLMs) are conscious (Chalmers 2023; Long 2023). This is because they appear to lack the architectural features described in our best theories of consciousness (e.g. information integration or a global workspace; Chalmers 2023).

### 5.1.2 Sources of uncertainty regarding AI consciousness

The current state of expert opinion might optimistically be termed a consensus. Even so, this concurrence is highly unstable at best. On the one hand, expert agreement is all but assured to be fractured with continued advancements in AI. On the other hand, there remains much uncertainty regarding several important aspects of AI consciousness:

i. **We don't really understand what consciousness is**. In other words, we lack a general theory of how consciousness arises from non-conscious matter and processes (e.g. neurochemical reactions)—we do not have a solution to the "hard problem" of consciousness (Chalmers 1995a). Although there exist different proposals (e.g. global workspace theory, integrated information theory, etc.; Hildt 2022; Ferrante et al. 2023), neither commands broad consensus among experts.

ii. **We don't have good ways to empirically test for consciousness in non-humans**. Partly as a result of the foregoing doubts regarding the very possibility of nonbiological consciousness, there currently exists no agreed-upon standard for testing for consciousness in AI (Dung 2023a; Chalmers 2023).These issues are discussed at length in the sections to follow.

iii. **We don't really understand how current AI systems (i.e., deep learning models) work**. Overarching "grand theories" of consciousness do not tend to enable specific predictions about which things are conscious (Dung 2023a). In any case, the physical mechanisms that give rise to consciousness in living things may not be the same as those that give rise to consciousness in machines (ibid). Properly diagnosing consciousness will likely require both "top-down" and "bottom-up" work. Among other things, this will require a deeper understanding of how various types of AI systems work (Saad and Bradley 2022).

iv. **"Risk of sudden synergy"**. It's hard to forecast future developments in AI consciousness because progress may not be linear. It is possible that a convergence of lines of research may lead to abrupt and exponential progress (Metzinger 2021a).

As it stands, we are in a state of radical uncertainty about future conscious AI (Metzinger 2021b). Continued progress in (i) consciousness research, (ii) consciousness evaluations (§5.1.3), and (iii) interpretability[39] may go some way towards mitigating this uncertainty. In the subsequent sections, we focus on (ii), touching on outstanding obstacles to reliable empirical tests of consciousness.

### 5.1.3 The current state of consciousness evals

The primary method of assessing consciousness in humans is through introspection and verbal report. However, this method is problematic in two respects: it is both over-exclusive and over-inclusive.

- **Over-*ex*clusive**: Introspection and verbal report cannot be used to ascertain consciousness in beings that are incapable of language, such as infants, individuals with various linguistic disabilities, or animals.
- **Over-*in*clusive**: The verbal reports of AIs such as current LLMs cannot be trusted as evidence of their consciousness, given that (1) their responses can be influenced by leading questions and (Berkowitz 2022) (2) the texts on which they have been trained include verbal reports of subjective experience produced by humans (Labossiere 2017; Andrews and Birch 2023; Chalmers 2023; Long 2023).

To be sure, dedicated tests of machine consciousness have a long history[40]. Despite this, there remains no widely accepted test of AI consciousness. Part of the reason for this is because, as previously mentioned (§2.2), there is still disagreement over whether it is possible for AI to be conscious in the first place. But this is not the only obstacle. Even if researchers agreed that a machine could, in principle, be conscious, the question of how to detect and diagnose consciousness would still remain[41]. In fact, many philosophers argue that consciousness is simply not the kind of thing that is amenable to empirical study (Jackson 1982; Robinson 1982; Levine 1983): that no amount of empirical information could *conclusively* establish whether or not a thing actually is conscious. In response to this concern, one might adopt a precautionary approach (§2.3.1)—rather than conditioning AI's moral status upon our knowing, beyond the shadow of a doubt, that they are conscious, or even *strong credence* (Chan 2011; Dung 2023a), we might settle for a weaker standard: a *non-negligible chance* (Sebo and Long 2023). This approach has been profitably pursued in animal welfare advocacy (Birch 2017).

As a result of the above doubts about the in-principle possibility of machine consciousness, as well as slow progress in AI development during the AI winters, progress on consciousness evals has largely been relegated to the field of animal sentience[42]. Indeed, conceptual and empirical research on animal sentience has often served as the template for research into AI consciousness (Dung 2023b; *c.f.* Tye 2017). Such efforts have proven to be of immense value in three respects:

---

[39] Interpretability work that may clarify the question of AI consciousness might include work on representation engineering (Zou et al. 2023) or digital neuroscience (Karnofsky 2022).
[40] For a review, see (Elamrani and Yampolskiy 2019). For a recent proposal, see (Schneider 2019).
[41] Just knowing that AIs could, in principle, be conscious still leaves us with the question of *which* AIs are conscious. In the field of animal sentience, this has been called the "distribution question" (Allen and Trestman 2024). Elamrani and Yampolskiy (2019) term this the "other minds" problem.
[42] In this usage, animal "sentience" essentially means the same thing as animal "consciousness". The usage of sentience indicates an emphasis on the ethical motivations behind animal welfare efforts, which go beyond disinterested "intellectual" interests.

i. **Conceptual precisification**: Tests of animal sentience have served to disambiguate the notoriously elusive notion of consciousness into more precise capacities. For instance, Birch et al. (2020) distinguish between five different dimensions: (i) self-consciousness/selfhood (the capacity to be aware of oneself as distinct from the external world), (ii) phenomenal richness (the capacity to draw fine-grained distinctions in a given sensory modality), (iii) evaluative richness (the capacity to undergo a wide range of positively or negatively valenced experiences), (iv) unity/synchronic integration (the capacity to bring together different aspects of cognition into a single, seamless perspective or field of awareness at a given point in time), and (v) temporal continuity/diachronic integration (the capacity to recall past experiences and simulate future experiences, all while relating them to the present).
This conceptual refinement has important consequences for discussions of AI consciousness. If consciousness can be decomposed into qualitatively distinct elements such as these, then it does not make sense to speak of different creatures as "more conscious" or "less conscious" (ibid; §2.4). Rather, they may better be thought of as "conscious in different ways" (Hildt 2022; *cf.* Coelho Mollo *forthcoming*).

ii. **Empirical tractability**: As a result of being better defined, the above-mentioned capacities proposed by Birch et al. promise to facilitate empirical testing. In the same paper, Birch et al. catalogue specific questions that have been explored under experimental paradigms under the foregoing five dimensions of consciousness (2020). These provide avenues for future diagnostics—including testing for consciousness in AIs (Dung 2023b). Satisfying multiple empirical measures of consciousness could provide defeasible grounds for believing that a given creature or AI is conscious (Ladak 2021).

iii. **The theory-light paradigm**: Theorists studying AI consciousness often draw a distinction between two broad approaches[43]. The "top-down" approach to AI consciousness involves first articulating a general theory of consciousness, and then drawing inferences from this to specific creatures or AIs (Dung 2023a). Over the years, this approach has been forestalled by chronic philosophical and scientific disagreements (Levine 1983). Discontent to defer animal welfare efforts to the fulfilment of this agenda, researchers are increasingly adopting a "bottom-up" (Birch et al. 2022) and "theory-light" (Birch 2020) approach to animal and AI consciousness[44]. This approach is defined by a minimal commitment to some sort of relationship between cognition and consciousness (ibid). There are no further overarching tenets, nor is any particular feature of consciousness held to be essential (Ladak 2021). In this way, the theory-light approach enables researchers to make empirical headway on questions related to welfare concerns without waiting for a general theory of consciousness.

---

[43] Elamrani and Yampolskiy (2019) differentiate between "architectural" and "behavioural" tests of AI consciousness. While the former places an emphasis on the structural and formal implementation of consciousness, the latter focuses on its outward manifestations. This corresponds roughly to Dung's (2023a) "top-down"/"bottom-up" dichotomy.

[44] Birch defines the theory-light approach in terms of a minimal commitment to what he calls the *facilitation hypothesis*: that "Phenomenally conscious perception of a stimulus facilitates, relative to unconscious perception, a cluster of cognitive abilities in relation to that stimulus" (2020). In other words, there is some kind of link between phenomenal consciousness and cognition. On this assumption, psychometric evaluations may be able to support inferences about subjective experience.
More recently, Chalmers (2023) proposes an alternative to Birch's theory-light approach, which he calls the "theory-balanced" approach. The theory-balanced approach assigns probabilities for a given thing's being conscious according to (1) how well-supported different theories are and (2) the degree to which that thing satisfies the criteria of these theories. At this point, neither the theory-light approach nor the theory-balanced approach appears to command an obvious advantage. As far as research in animal and AI consciousness is concerned, the upshot is that, today, there are serious contenders to the traditional "theory-heavy" approach which commit to particular theories of consciousness, enabling multiple paradigms to be developed in parallel.

## 5.1.4 Prevailing challenges to testing for consciousness in AI

Today, there is a new wave of interest in AI consciousness evals[45] (Schneider 2019; Berg et al. 2024a; Sebo 2024). Nonetheless, much work is still needed in order to devise reliable empirical tests of AI consciousness. Four major outstanding issues include:

i. **How to make increasingly fine-grained assessments of specific symptoms or criteria of consciousness?** With continued research, consciousness as we know it in the broad sense is increasingly refined into narrower, better defined capacities (Birch et al. 2022). These more granular capacities can themselves be assessed in even more precise ways. Take, for instance, the capacity of phenomenal richness ("*P*-richness"): the capacity to draw fine-grained distinctions in a given sensory modality (e.g. vision, olfaction…)[46,47]. Birch et al. (ibid) note that *P*-richness can itself be further resolved into *bandwidth* (e.g. the amount of visual content that can be perceived at any given moment), *acuity* (the fineness of "just-noticeable differences" which can be detected), and *categorisation power* (the ability to organise perceptual properties under high-level categories[48]). Whether and how these, in turn, may be refined into subtler, more empirically robust capacities remains to be seen.

ii. **How to integrate different symptoms or criteria into an overall assessment of consciousness?** The increasing proliferation of empirical measures of consciousness under the open-ended theory-light paradigm raises the question of how to bring these qualitatively distinct criteria together into a holistic evaluation. Mature tests of AI consciousness will need to systematically negotiate these diverse factors—for instance, by weighing different criteria according to Bayesian rules[49] (Muelhauser 2018; Ladak 2021), or exploring the possibility that certain combinations of traits may be synergistic (or even antagonistic). Importantly, this question will become increasingly critical as empirical measures of consciousness become increasingly refined (see above discussion of *P*-richness).

iii. **How to account for AIs possibly gaming consciousness evals?** AIs present two unique challenges to consciousness evals (or evals of any sort, for that matter; Hendrycks *forthcoming*). First, they can be purpose-built by developers to pass or fail consciousness evals (Schwitzgebel 2024). Developers may harbour any number of motivations (financial or otherwise) to ensure that their AIs achieve certain results on consciousness evals, and can design models to behave accordingly in test scenarios. For instance, a lab seeking to skirt regulatory protections for conscious AI might prevent their models from being able to pass consciousness evals. Second, AIs themselves, if sufficiently situationally aware, might determine that certain results on consciousness evals are either conducive or detrimental to their goals. Andrews and Birch (2023) call this the "gaming" problem[50]. For example, in the false positive scenario, society might provide certain legal protections to AI that is "provably" conscious. Non-conscious AI that recognises the strategic value of these benefits with respect to its own goals may aim to pass consciousness evals (or design future systems capable of passing consciousness evals). In the long run, this can lead to human disempowerment ([§4.2)](#).

---

[45] At the time of writing, Jeff Sebo and Robert Long are currently working on a consciousness eval for AI: a standard test which draws upon the latest research across both AI and animal sentience (Sebo 2024).

[46] e.g. the ability to distinguish between two colour hues, subtle notes within a wine or perfume, or different textures. Research on "just-noticeable differences" (JNDs; Tabakov et al. 2021) in experimental psychology attempts to pinpoint perceptual thresholds of stimulus intensity.

[47] Since creatures may differ in the sensory modalities they possess (e.g. humans lack echolocation, robots can be built without the capacity for taste), it does not really make sense to speak of a creature's "overall" level of P-richness (Birch et al. 2022). Nor, of course, does the absence of a given sensory modality necessarily discount the creature's likelihood of being conscious.

[48] Lexical concepts such as colour terms ("beige", "eggshell") or words for flavours ("salty", "sweet").

[49] Dung (2023a) advances a structured deliberative process for assessing AI consciousness. Fischer's "moral weights" approach (2022; 2024) provides an alternative paradigm centred on welfare rather than consciousness.

All this shows that a variety of approaches will have to be combined in order to ensure the reliability of AI consciousness evals (Hildt 2022). Among other things, this includes research on "negative criteria"—or defeaters: features which count *against* an AI's being conscious. One simple example of this is an AI's having been designed specifically to exhibit certain features of consciousness. In such cases, the AI's seeming to be conscious is simply *ad hoc* (Dung 2023b) or gerrymandered (Shevlin 2020; *cf*. Schwitzgebel 2023) as opposed to "natural".

AI consciousness evals might become increasingly sensitive and specific, reducing the risks of false positives and false negatives. However, beyond a certain point, further increasing one often comes at the expense of the other due to inherent trade-offs in test design (Doan 2005). Precautionary motivations [(§2.3.1)](#) favour tolerating more false positives. That being said, the diagnostic biases of consciousness evals will likely need to be adapted in response to evolving technological, cultural, and social conditions.

iv. **Where is this all going?** It bears emphasis that the theory-light approach is a stopgap tactic. For the time being, the theory-light approach performs the crucial function of allowing conceptual and empirical consciousness research programmes to progress in parallel—all in the service of potential moral patients [(§2.3.1)](#). In the long run, however, it will become increasingly necessary to build explicit and substantive connections between theory and measurement. Firstly, such connections may help mitigate concerns about AI gaming consciousness evals (Hildt 2022). Secondly, merely suspending theory does not amount to absolute theoretical impartiality. Researchers operating under the theory-light approach might nonetheless unwittingly embed assumptions about consciousness (Thagard 2009; *cf*. Kuhn 1970). What is *nominally* "theory-light" may later turn out not to be so neutral after all. Disastrous consequences can follow if implicit, unreflective assumptions about consciousness are allowed to influence policy[51]. Thirdly, concepts of consciousness might drift apart and eventually diverge into fully distinct notions (e.g. biological consciousness vs. machine consciousness; Blackshaw 2023). While natural conceptual evolution is not inherently problematic, this can become an issue if this divergence leads to objectionably preferential treatment. This can manifest in a future scenario in which AIs are considered "conscious but not in the sense that matters" (e.g. in a biological sense). In this case, a bioessentialist [(§2.2)](#) concept of consciousness would, all else being equal, prioritise the interests of living things to those of AIs. The point is not that a unified concept of consciousness ought to be preserved at all costs. Rather, it is that all stakeholders should be sensitive to the material consequences of conceptual drift and revision (Haslanger 2000).

---

[50] The gaming problem need not require AI to have genuine deceptive intentions. Nonetheless, it is also conceivable that sufficiently intelligent AI could come to harbour deceptive intentions, and conduct itself differently under testing conditions—including consciousness evals.

[51] In particular, such assumptions could impact which AIs receive moral consideration, and the degree to which their interests are taken into account.

## 5.2 Epistemic axis

Right now, AI consciousness receives relatively little attention compared to other AI-related concerns, such as bias and discrimination, misinformation and disinformation, worker displacement, and intellectual property infringement (Google Trends *n.d.*; §1.1). Nevertheless, when asked about it, most people do express some level of concern about AI consciousness—the modal view today is that it is possible for future AIs to be conscious, and if they were, they should be given some degree of moral consideration.

### 5.2.1 What does the general public believe about AI consciousness today?

A 2023 poll by the Verge and Vox Media's Insights and Research team of 2,000+ American adults found that around half "expect that a sentient AI will emerge at some point in the future" (Kastrenakes and Vincent 2023). Similarly, according to the 2023 Artificial Intelligence, Morality, and Sentience (AIMS) survey conducted by the Sentience Institute (Pauketat et al. 2023), nearly 40% of Americans believe it is possible to develop sentient AI, compared to less than a quarter of American adults who believe it is impossible. Some proportion of individuals—5% according to a poll by Public First (Dupont et al. 2023) and 19% according to the AIMS survey—even believe that some of the AIs we have today are already sentient. However, it is worth highlighting that these polls focus on the US and UK—there is currently a lack of data on the views of other regions, which AI agents will be also subject to the laws and treatment of.

As with most other topics, it seems likely that these beliefs will vary between different demographics. Age and gender have already been shown to influence people's opinions on other AI-related topics—for example, young people and men are more likely to trust AI, while women and older people are less likely to trust it (Yigitcanlar et al. 2022). However, research into which demographics are more likely to believe AI consciousness is possible, whether it would be deserving of moral consideration if so, and which policies should be enacted to mitigate the possible harms is either scant or lacking.

Moreover, there does seem to be a level of agreement among the public that sentient AIs would deserve some degree of moral consideration, though short of being treated equally to humans. This moral consideration translates to varying degrees of support for different public policies (figures 1 and 2, *over*).

Increasing moral consideration for conscious AIs also seems to be a tractable issue. Lima et al. (2020) have shown that certain interventions can have "remarkable" and "significant" effects in promoting support for AI sentience policies. Namely, showing respondents examples of non-human entities that are currently granted legal personhood (e.g. corporations or, in some countries, nature) shifted the opinion of 16.6% of individuals who previously were opposed to granting AIs legal personhood. Additionally, Allen and Caviola (2024) showed that having a brief conversation with an LLM "loaded with a prompt to persuade users that harming [digital humans] would be as bad as harming a typical human" significantly increased participants' level of moral concern. This shows that shifting public opinion on this topic may be tractable through educational campaigns.

That said, it is worth highlighting that even if people support AI welfare in principle, it is unclear to what extent this will translate into a willingness to make real trade-offs in the name of AI welfare. Using animal welfare as an analogue, many people believe farmed animals are sentient and, in theory, would prefer for them not to be subjected to factory farm conditions—but in practice, few are willing to stop buying animal products or make other lifestyle changes to prevent harms to animals (Ladak and Anthis 2022). Additionally, the figures above are subject to considerable uncertainty—in the past, a similar survey on factory farming by the Sentience Institute has failed to replicate (Dullaghan 2022).

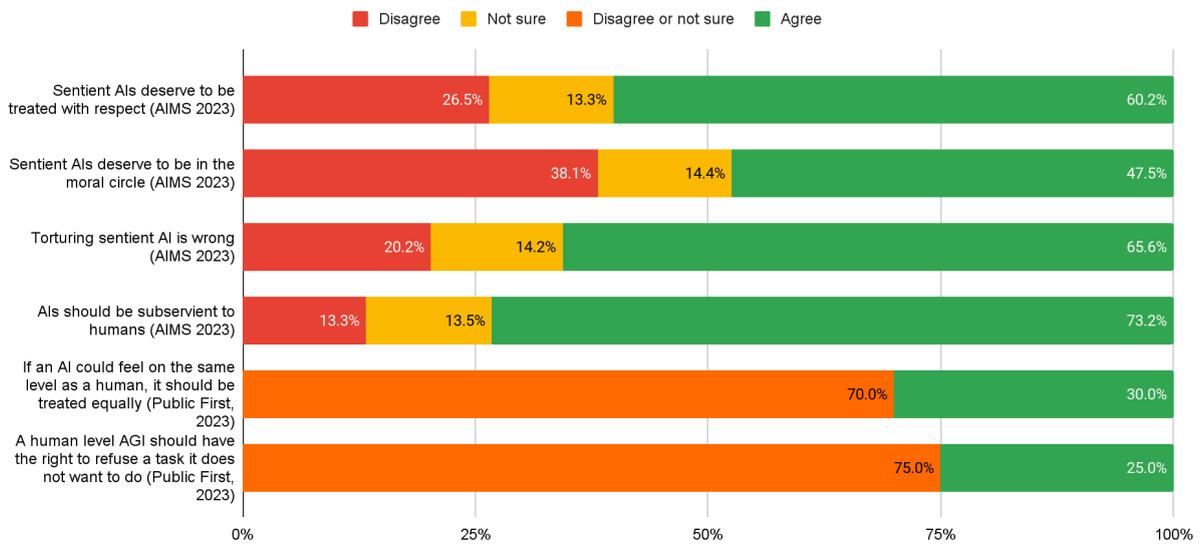

Figure 1, **GENERAL PUBLIC'S MORAL VIEWS REGARDING SENTIENT AI**: Data from (Pauketat et al. 2023).

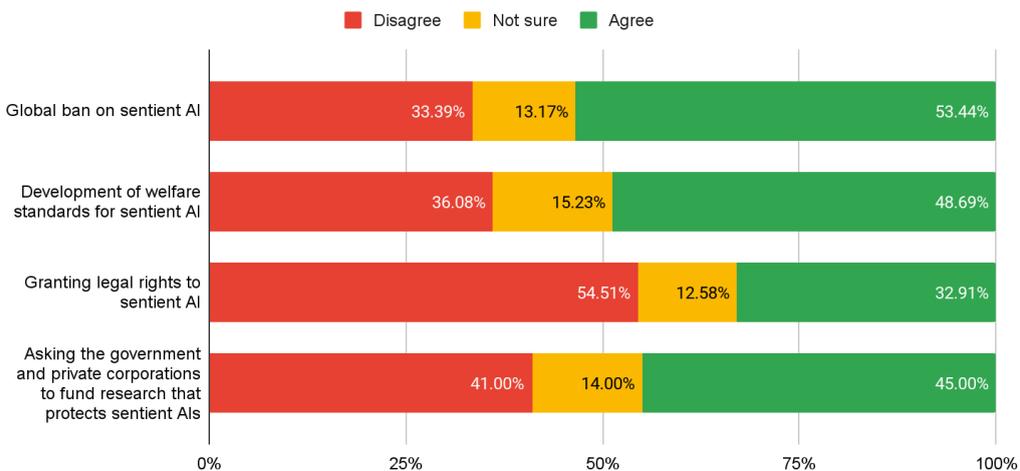

Figure 2, **GENERAL PUBLIC'S POLICY VIEWS REGARDING SENTIENT AI**: Data from (Pauketat et al. 2023).

### 5.2.2 How are the public's beliefs about AI consciousness likely to evolve?

It seems likely that as AI becomes more advanced and anthropomorphic, the proportion of people that believe it is sentient will increase. Vuong et al. (2023) found that "interacting with an AI agent with human-like physical features is positively associated with the belief in that AI agent's capability of experiencing emotional pain and pleasure" (see also Perez-Osorio and Wykowska 2020). Similarly, Harris and Anthis (2024) tested the effect of 11 features on moral consideration for AI, and found that "human-like physical bodies and prosociality (i.e., emotion expression, emotion recognition, cooperation, and moral judgment)" had the greatest impact in people's belief that harming AI is morally wrong.

Anecdotally, the anthropomorphization of AI seems to be already materialising—some users of AI companion apps like Replika have come to believe their AI partners are sentient (Pugh 2021; Dave 2022). These relationships may become increasingly common, and contribute to the belief that AIs can experience feelings—in a poll conducted by the Verge, 56% of respondents agreed that "people will develop emotional relationships with AI", and 35% said they "would be open to doing so if they were lonely".

On the other hand, there might also be reasons for the opposite to occur, i.e. anthropodenial (de Waal 1997). For instance, if treating conscious AIs as tools would be unethical, people would have an incentive not to believe AIs are conscious. This would be precedented by the anthropodenial humanity has engaged in towards other animal species, or historically, towards humans of different ethnicities (Lifshin 2022). It is also worth highlighting that the anthropomorphising of AIs may be harmful, as the way AIs prefer to be treated might differ from the way humans prefer to be treated—Mota-Rojas et al. (2021) have outlined how a similar dynamic plays out in interactions towards companion animals.

Another relevant consideration is the influence of experts' views on the public's views. There is evidence that these two groups think about conscious experience differently—namely, Sytsma and Machery (2010) found that, compared to philosophers, laypeople were more likely to attribute phenomenal consciousness (e.g. the ability to see red) to robots while rejecting the idea that they felt pain. At the same time, a survey conducted by Caviola (2024a) shows the views of experts and the general public on some questions akin to those above are similar (figure 3, *below* and figure 4, *over*).

In the areas where they diverge, it is unclear what effect expert opinion will have on the general public's beliefs. On other topics, such as climate change and vaccines, we have seen the public fracture along 'high trust' and 'low trust', and place different weight on the views of experts accordingly (Kennedy and Tyson 2023). However, given the philosophical nature of this topic, it seems possible that people will be especially likely to dismiss 'expert opinion'. Davoodi and Lombrozo (2022) have shown that people distinguish between scientific unknowns (e.g. *What explains the movement of the tides?*) and "universal mysteries" (*Does God exist?*). It is possible that many people do not consider consciousness a scientifically tractable topic, and will consider the views of experts less relevant than their own intuition for this reason.

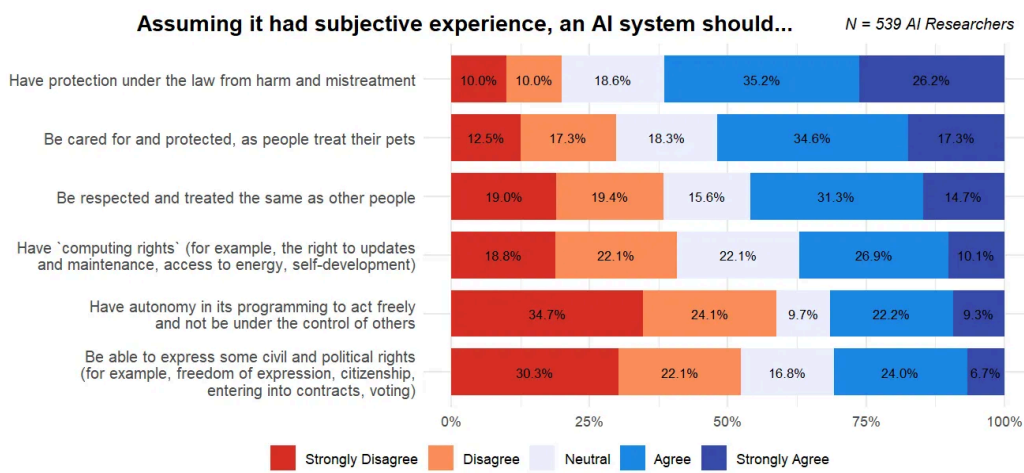

Figure 3, **EXPERTS' POLICY VIEWS REGARDING SENTIENT AI**: Figure reproduced from (Caviola 2024a).

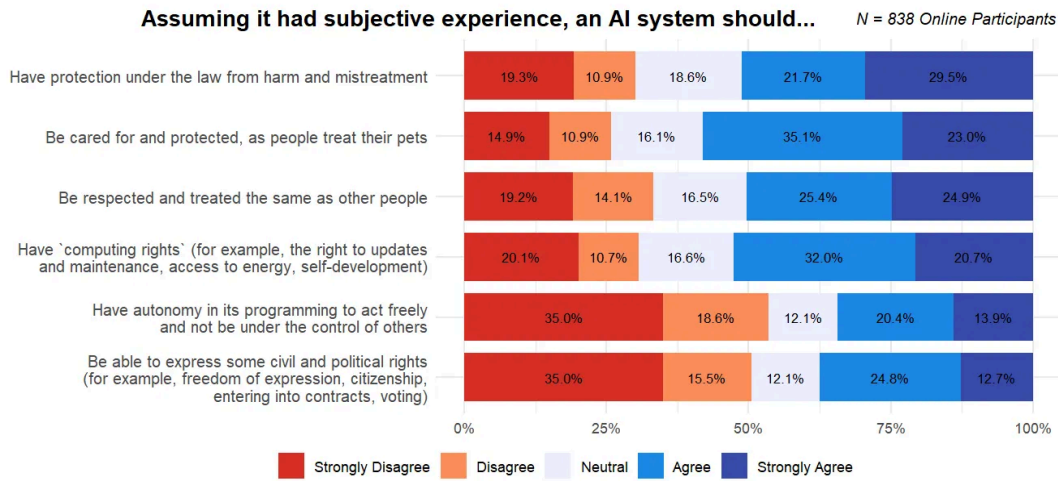

Figure 4, **GENERAL PUBLIC'S POLICY VIEWS REGARDING SENTIENT AI**: Figure reproduced from (Caviola 2024a).

# 6. Recommendations

## 6.1 Review of major risks related to conscious AI

In (§4.2), we discussed four major risks related to conscious AI: AI suffering, human disempowerment, depravity, and geopolitical instability. In this section, we propose interventions aimed at reducing these risks. To summarise our proposals:

- **Recommendation I**: We recommend taking measures to avoid intentionally building conscious AI.
- **Recommendation II**: We support research in the following areas: consciousness evals, how to enact AI welfare protections in policy, surveys of expert opinion, surveys of public opinion, and provisional legal protections for AI.
- **Recommendation III**: We support the creation of an AI public education campaign (specifically related to conscious AI).

Table 5 (*over*) summarises how each proposed recommendation targets the aforementioned risks.

| Recommendations ↓ | Risks → | AI suffering | Human disempowerment | Depravity | Geopolitical instability |
|---|---|---|---|---|---|
| I. **Don't intentionally build conscious AI** | | ✓ | ✓ | | ✓ |
| II. **Supported research** | *Consciousness evals* | ✓ | | | |
| | *AI welfare policy* | ✓ | | | |
| | *Expert surveys* | ✓ | | | |
| | *Public surveys* | | ✓ | ✓ | ✓ |
| | *Provisional legal protections* | ✓ | | | |
| III. **AI public education campaigns** | | ✓ | ✓ | ✓ | ✓ |

Table 5, **RISKS AND RECOMMENDATIONS**: This table depicts how each of our recommendations addresses the aforementioned risks.

Before delving into the recommendations, we briefly comment on how exactly we evaluate different options for interventions.

## 6.2 What makes a good intervention?

We aim to take a principled approach to determining the best and most effective strategies for reducing risks related to conscious AI. To this end, we shall adopt the four following criteria to assess the advantages and disadvantages of possible interventions[52] (Dung 2023a):

1. **Beneficence**: The proposed intervention should reduce net suffering risk (of humans, machines, animals, and other moral patients) by as much as possible. At the same time, it should not significantly increase the probability/severity of other negative outcomes nor decrease the probability/utility of other positive outcomes.

2. **Action-guiding**: The proposed intervention should suggest sufficiently concrete courses of action. That is to say, there is minimal vagueness in how it ought to be implemented. This also means that interventions should not imply conflicting or contradictory courses of action.

---

[52] These 4 criteria partially coincide with 3 major risk aversion strategies from decision theory applied in animal sentience research (Fischer 2024). (1) *Worst-case aversion* prioritises minimising the potential for highly negative outcomes. Options are ranked based on perceived severity of the worst possible outcome rather than solely on expected value maximisation. This strategy optimises for security and safety, and may coincide with a focus on beneficence. (2) *Expected value maximisation* prioritises interventions that are more likely to produce concrete, measurable positive results. This approach optimises for measurability and effectiveness, and may coincide with a focus on feasibility. (3) *Ambiguity aversion* prioritises interventions that rely the least upon uncertain information. Options are ranked based on estimated proportion of knowledge vs. ignorance of the key facts. This means that interventions that rely upon unlikely but known probabilities are preferred over interventions that rely upon assumptions with a wider probability distribution. This strategy optimises for clarity and confidence in decision-making processes, and may coincide with a focus on action-guiding and feasibility.

3. **Consistent with our epistemic situation**: The proposed intervention should broadly presuppose uncertainty regarding prevailing theoretical issues, the future trajectory of technological advances, and future sociocultural developments. In other words, interventions should not require knowledge that is beyond our current grasp and especially knowledge that we are unlikely to attain in the immediately foreseeable future (e.g. knowledge of how the capacity for phenomenal consciousness can arise from non-conscious matter).

4. **Feasibility**: The proposed intervention should be realistic to implement. In other words, for the relevant actor[53] (i.e. an AI lab or governmental agency), the intervention should not only be practically possible to implement, but there should also be a reasonable expectation of success.

The order of our recommendations reflects our overall judgments with respect to these criteria.

## 6.3 Recommendation I: Don't intentionally build conscious AI

### Description

- **Overview**: Our first recommendation is to pursue regulatory options for halting or at least decelerating research and development that directly aims at building conscious AI.

- **Justification**: Attempts to build conscious AI court multiple serious risks (AI suffering, human disempowerment, and geopolitical instability; §4)—endangering not only humans but also AI[54]. What's more, as we argue in appendix (§A1.1), the positive motivations for building conscious AI (improved functionality, safety, insights into consciousness) are nebulous and speculative. Both experts (Saad and Bradley 2022; Seth 2023) and the general public[55] (§5.2.1) support the idea of stopping or slowing efforts to build conscious AI.

- **Conditions of implementation**: Perhaps the most specific and well-discussed proposal comes from the philosopher of mind Thomas Metzinger (2021a), who has also served on the European Commission's High-Level Expert Group on AI[56]. Metzinger's proposal consists in two parts[57]:

    i. A global moratorium on lines of research explicitly aimed at building conscious AI, lasting until 2050 and open to amendment or repeal earlier than 2050 in light of new knowledge or developments

    ii. Increased investment into neglected research streams

---

[53] Most interventions will require the action of other parties besides oneself. In such cases, feasibility should not presuppose their backing of the intervention. Rather, feasibility will also need to take into account the relative probability that other necessary actors can be convinced to enact the desired intervention—in addition to the probability that, if enacted, the intervention will achieve positive results (Dung 2023a).

[54] Consistent with a preventative ethics approach (Seth 2021), by avoiding building conscious AI, we avoid bringing vulnerable individuals into existence. By the same stroke, we also sidestep future moral dilemmas related to conscious AI, including such thorny issues as trade-offs between the welfare of humans and that of AIs, as well as how to live alongside morally autonomous AI (ibid). Conscious AI is a genie that, once unleashed, is difficult to put back into the bottle. That is to say, it will be difficult to revert back to a state before conscious AI without committing some kind of violence.

[55] The 2023 Artificial Intelligence, Morality, and Sentience (AIMS) survey found that 61.5% of US adults supported banning the development of sentient AI (an increase from 57.7% in 2021; Pauketat et al. 2023).

[56] In (2019), the High-Level Expert Group published *Ethics Guidelines for Trustworthy AI*. However, Metzinger has later criticised the guidelines (2019; 2021b). On his view, the guidelines were industry-dominated (suffered from "regulatory capture"), short-sighted (largely ignored long-term risks including those related to conscious AI and artificial general intelligence), and toothless (adherence is not compelled by enforceable measures).

[57] Both aspects of this proposal can be dated back to the Effective Altruism Foundation's 2015 policy paper on AI risks and opportunities, which Metzinger co-authored (Mannino et al. 2015).

Our discussion of recommendation II (§6.4) expands on (ii). As concerns (i), the main issue is determining which lines of research are affected. Currently, multiple actors across academia and industry openly aim at building conscious AI. One place to start could be for governments and funding institutions to restrict funding to projects whose explicit goals are to build conscious AI. Second, research projects whose goals are of strategic relevance to conscious AI (and, in particular, AI suffering) may also fall under the scope of this moratorium. This could include, for instance, work on creating AI with self-awareness (self-models) or affective valence.

This selective scope improves the feasibility of this intervention relative to global bans[58] on AI research (Dung 2023a). Furthermore, because the benefits of conscious AI are unclear (§A1.1), we believe a ban on conscious AI research is unlikely to trigger a "race to the bottom" (Tegmark 2023; though see §3.3.2 on rogue actors building conscious AI).

## Special considerations

- **Action-guidance:** Various details of Metzinger's proposal are subject to debate. These include: (1) when would be the most opportune moment to initiate the ban, (2) for how long should the ban be sustained to be most effective, and (3) how best to operationalise the scope of the ban? We do not attempt to resolve these issues in this policy paper. We note them as key priorities for future work. Having said that, future work on these questions may profit from existing debates over pause AI, which have yielded helpful and constructive frameworks for further discussion (see especially Alexander 2023). Finally, other key details to be resolved include the conditions for amendment or repeal: what new developments or achievements could warrant: (a) revising the scope of the moratorium, (b) terminating it before 2050, or (c) extending its duration past 2050? These details are also critical to the overall promise of this intervention.

## 6.4 Recommendation II: Supported research

### 6.4.1 Consciousness evals

#### Description

- **Overview**: We still lack a robust way to test for the presence of conscious experience in artificial systems (§5.1.3). Further work could take the direction of the theory-light approach, borrowed from animal welfare research. Specifically, we propose focusing on the following questions:

    1. What capacities should be assessed? How to define and refine them so that they are empirically tractable and robust?
    2. How to integrate the different criteria into an overall assessment?

  Researchers could also investigate other avenues like the recent effort with self-reports and interpretability (Perez and Long 2023). Efforts should be aimed at reducing the risk of false negatives and false positives, and devising tests which cannot be (easily) gamed.

- **Justification**: Even if we are successful in limiting direct efforts to build conscious AIs, consciousness might still emerge as an unexpected capacity. We need to be able to say whether a system is conscious, since this will inform whether this system could be treated as an object or not. We should be able to do so with higher certainty than we do now. While a complete theory of consciousness would be very helpful, we could still make progress towards evaluating the moral significance of AI systems by developing tests for consciousness, which could at least reduce our confidence intervals.

---

[58] The overall pace of AI progress (whether or not related to conscious AI) is a major concern among experts (Grace et al. 2024). Movements such as "Pause AI" or "effective decelerationism" advocate for global restraints on AI progress. Having said that, these conservative movements are not without their critics (Lecun 2024). Outright global bans on all AI research are likely infeasible due to the economically and geopolitically strategic value of advanced AI capabilities, and due to the multilateral coordination that would be required for such a ban to be effective (Dung 2023a).

Special considerations

- **Consistency with our epistemic situation**: Creating tests for consciousness should be done with future technological advances in mind, since progress in AI systems might make certain tests obsolete and/or possible to game.

### 6.4.2 AI welfare research and policies

Description

- **Overview**: If future AI systems merit moral consideration, we need to understand what interests they'll have and how to protect them (Ziesche and Yampolskiy 2019; Hildt 2022). AI welfare research should investigate the former, while devising AI rights should address the latter.
Exploration of AI welfare policies depends on the kinds of moral status artificial systems will have. Legal protections will be different for AI systems with a moral status akin to some animals, where degrees of certainty w/r/t their consciousness vary for different species, compared to for systems with status akin to human-level status, if not higher (Shulman and Bostrom 2021). Therefore, a key question to answer is about the types and degrees of moral status different AI systems will have, which might depend on the qualities and degrees of their consciousness (Hildt 2022). Similarly to assigning moral status to animals based on taxonomy, we could look for ways to categorise AI systems based on architectural or functional features. We therefore see AI welfare research to develop as a specific line of artificial consciousness research (Mannino et al. 2015).
When it comes to AI welfare policies, it is important to note that AI rights need not necessarily be against human rights. We should strive to find positive-sum solutions where possible—actions which are beneficial for both humans and AIs (Sebo and Long 2023); and be open to compromises if they can have significant impact (like making changes to one's diet for environmental or animal welfare considerations).
- **Justification**: Even if future AI systems are probably not conscious, it is worth exploring what it might mean for future AI systems to be treated well, so that we are prepared for this possibility. Similarly, even in the face of current uncertainty around the kinds of moral status they'll have, we can start thinking about policies under different scenarios.
- **Conditions of implementation**: These efforts require highly interdisciplinary work. Consciousness researchers should work together with AI developers to define what welfare means for AI systems, and together with ethicists to resolve ambiguities around their moral status. This work will feed into policy efforts to find positive interventions for both humans and nonhumans.

Special considerations

- **Uncertainties**: Assessing AI welfare and how to improve it might not be feasible until we have a more mature theoretical understanding of consciousness. More importantly, certain directions for policies might turn out to be misguided and even potentially harmful.
- **Alternative frameworks for AI rights**: While the most conventional reasoning for AI protections is pathocentrism, it is worth noting that some scholars encourage alternative frameworks, such as the social-relational framework which "grants moral consideration on the basis of how an entity is treated in actual social situations and circumstances" (Harris and Anthis 2021; Gunkel 2022). It might be that we should give AI systems rights even if they are not conscious (§6.4.6). In this case, the discussion of AI rights should take into account not only consciousness evals, but also the aims and needs of the broader societies into which AIs are integrated.

### 6.4.3 Survey on expert opinion

#### Description

- **Overview**: Taking into consideration a recent academic survey on consciousness (Francken et al. 2022) of researchers with a variety of backgrounds (e.g. philosophy, neuroscience, psychology, and computer science), as well as a large-scale survey of AI practitioners (Grace et al. 2024) on the future of AI, we propose devising a large-scale survey on machine consciousness which will combine aspects of both:
    a. Respondents will be presented with a number of questions about the nature of artificial consciousness. For example, they could be presented with a list of conditions for consciousness and be asked to assign a probability for each to be necessary or sufficient, as well as a probability for whether it would be present in AI systems by a certain year (this kind of analysis is done by Sebo and Long (2023))
    b. Respondents will show what levels of concern they have about different negative scenarios and risks related to conscious AI.

- **Justification**: Conducting a survey on expert opinion on AI consciousness will be useful for the following reasons:
    a. It can inform work on consciousness evals.
    b. It can identify the kinds of AI work which might be indirectly focused on building conscious AI (and this might be relevant for the implementation of a potential ban on building conscious AI, see [§6.3](#)).
    c. It can identify major uncertainties or neglected areas and therefore where to focus future research.
    d. It can help prioritisation of work on certain risks over others.
    e. Repeating the survey on an annual basis will allow monitoring trends.

- **Conditions of implementation**: It will be valuable to include a wide range of researchers—AI experts and consciousness experts. Undoubtedly, the most informed answers will be from experts whose focus is artificial consciousness. However, there may not be many of them (in Francken et al's (2022) survey, only 10% of the total 166 respondents had a background in computer science) and in addition, there will be value in having a diverse set of views. The risk of uninformed views is higher for the AI practitioners than for consciousness practitioners since the overlap between consciousness *in general* and *artificial* consciousness is bigger than between AI and conscious AI. The idea is that consciousness experts will be better able to comment on AI consciousness than computer scientists. However, there is a considerable benefit to including different AI specialisations since it is possible that some AI developers, despite not being *directly* involved with conscious AI, are indirectly working towards it by working on capabilities which might be conditions for AI consciousness.

#### Special considerations

- **Questions regarding moral status**: Introducing assessments of the severity of a number of scenarios and risks might entail giving assessments for the moral status/weight of future AIs. In other words, experts might implicitly make judgments about the comparative value of AIs with respect to humans. Schukraft (2020) argues that this method for measuring capacity for welfare status can be misleading since moral intuitions do not necessarily point to objective truths. This is why the questions which involve scenarios and risks related to conscious AIs should be formulated with care and should not be used to draw conclusions about the different degrees of moral status AIs should have.

### 6.4.4 Survey on public opinion

Description

- **Overview**: The data we currently have on the public's views about AI consciousness is limited. Therefore, we recommend the following:
    a. More public opinion polls about AI consciousness should be conducted, in general. Other than the AIMS survey (Pauketat et al. 2023), we were only able to find two other polls from recent years about this topic.
    b. Public opinion polls about AI consciousness should be conducted in countries other than the US and UK—currently, all publicly available polls on this issue are from these two countries. While these two countries will likely have an outsized impact on the development and legislation of AI consciousness, they only comprise around 5% of the global population combined—it is likely that many if not most future AI agents will be subject to the laws and treatment of the remaining 95%, making understanding their beliefs and attitudes important.
    c. Public opinion polls about AI consciousness should include demographic analyses, such as age, gender, religious affiliation, political orientation, and socioeconomic status, in order to examine how answers to other questions vary with these factors.
    d. Public opinion polls about AI should include questions designed to gauge how much respondents care about this issue, such as:
        i. How they would rank the importance of AI suffering relative to other AI risks (such as AI-generated misinformation or algorithmic bias), or how they would rank the importance of AI risks relative to other political issues.
        ii. How many government resources they think should be spent on risks related to AI suffering.
        iii. Whether they would be willing to make hypothetical tradeoffs to prevent AI suffering (e.g. pay $X a month more for a more humane LLM, analogous to the price gap between eggs from free-range vs. caged chickens).
- **Justification**: Among other things, this information would help forecast how future discourse about AI consciousness is likely to play out (for instance, whether it is at risk of being politically polarised) and would help determine how many resources to invest in interventions to sway public opinion (§6.5).
- **Conditions of implementation**: We would be keen to see independent researchers, think tanks, nonprofits, and media outlets alike conduct polls on these issues. These polls would be most effective if conducted on a periodic basis (e.g. annually) to monitor how opinions evolve.

Special considerations

- While these polls could show policymakers that their constituents care about this issue, they could also show the opposite. Therefore, there is a risk that it would cause a decrease in the amount of attention and public resources devoted to this problem. Pollsters should be mindful of this risk when publishing and discussing their results.

## 6.4.5 Decoding consciousness attribution

Description

- **Overview**: We also support research that is aimed at deciphering the logic behind people's intuitive consciousness attributions[59]: what factors drive people to perceive AI *as* conscious?

- **Justification**: Consciousness is a complex, multifaceted phenomenon. This being the case, a holistic understanding of consciousness must reconcile philosophical and scientific explanations with a thorough account of the factors driving consciousness perceptions[60]. This is crucial because consciousness attributions may be subject to *confounders*: factors which compromise the reliability of our judgments about which things are conscious. Such confounders might include cognitive biases underwriting anthropomorphic tendencies (biasing towards false positive judgments), or financial incentives to eschew regulations protecting conscious AIs[61] (biasing towards false negative judgments). Confounders may unwittingly prejudice consciousness attributions one way or the other, or they may be exploited by bad actors.
Research on the mechanics of consciousness attributions can reduce uncertainty about our intuitive perceptions of consciousness, promote responsible AI design (Schwitzgebel 2024), and help to steer away from the false negative and false positive scenarios.

- **Conditions of implementation**: Specifically, we propose two research directions:

    i. *Anthropomorphism and the intentional stance*:

        a. Which properties of AIs (e.g. anthropomorphic design cues; Perez-Osorio and Wykowska 2020) tend to elicit consciousness attributions from humans?

        b. What factors lead people to treat things as intentional agents (Dennett 1971, 1987)? What factors engage people's theory of mind reasoning capacities?

        c. What is the relationship between the uncanny valley effect[62] and consciousness perceptions? Given that the uncanny valley effect *can* be overcome through habituation (Złotowski et al. 2015), can certain interventions also positively influence consciousness perceptions?

    ii. *Relations with AI*:

        a. What *non*-technical[63] features or capacities (e.g. emotional intelligence) favour trustworthy perceptions of AI?

        b. How do macro-level factors (e.g. different actors' vested interests, cultural effects) influence their consciousness attributions? How to cultivate sensitivity to biases towards anthropomorphism and anthropodenial (de Waal 1997; Sebo and Long 2023)?

---

[59] Much like how people's tendency to attribute gender to different individuals tell us something interesting about their underlying concepts of gender (e.g. relative importance of biological factors, psychological identity, social roles…; Haslanger 2000), people's tendency to attribute *consciousness* to different things (AIs, animals, plants, rocks…) also tells us something interesting about their underlying concepts of consciousness (the relative importance of material constitution, cognitive sophistication, behavioural complexity…).

[60] In other words, the "scientific" and "manifest" images (Sellars 1963) of consciousness must be brought to terms with one another.

[61] e.g. dependence upon AIs to fulfil critical roles that would otherwise be disrupted by the recognition that they are conscious moral subjects

[62] The uncanny valley effect (Mori 2012) refers to a sensation of discomfort or unease that individuals feel when confronted with an artificial thing that appears almost, but not quite human.

[63] The object of interest here is perceived trustworthiness, not "actual" trustworthiness. Accordingly, explainability and interpretability are not relevant.

## 6.4.6 Provisional legal protections

### Description

- **Overview**: We support research which (1) clarifies the extent to which abusive behaviours towards AI can negatively influence relations towards humans and (2) in the event that this hypothesis is supported, explores strategies that could mitigate adverse spillover effects.

- **Justification**: Our final research recommendation addresses the risk of depravity [(§4.4)](). The risk of depravity essentially rests on an empirical hypothesis: that abusing life-like AI (especially social actor AI) might have detrimental effects on our moral character, eventually altering the way we behave towards other people for the worse. Although the specific mechanisms underpinning this transitive effect ("spillover") remain to be rigorously worked out, there is enough evidence of such a connection to warrant precautionary efforts (Guingrich and Graziano 2024). Moreover, the fact that depravity is a risk that occurs across all quadrants implies that precautionary efforts could be robustly efficacious despite uncertainty about which scenario we are in or will be in.

- **Conditions of implementation**: First and foremost, we suggest the following research directions aimed at testing and clarifying the transitive hypothesis:

  a. Are certain types of relations with AI (e.g. emotional, sexual, employee) more likely than others to have spillover effects on human relations? What sorts of factors promote or hinder this transitive effect (e.g. anthropomorphic design cues)?

  b. Which cognitive mechanisms subserve spillover effects?

  c. What effects might culture have on the type and extent of spillover effects?

  d. Could spillover effects be positively leveraged to support the cultivation of human virtues? In other words—could AI help us to treat other humans better, and, in doing so, become better people (Guingrich and Graziano 2024)?

If the transitive hypothesis fails to be supported by further research, then the risk of depravity can be safely dismissed. However, if evidence continues to mount in favour of the transitive hypothesis, then it would be prudent to explore policy options to prevent negative spillover effects resulting in harm to humans. One forthcoming strategy could be extending provisional rights and legal protections to AI. These concessions would be provisional in the sense that they do not depend upon strong proof that AIs are conscious (e.g. passing a standardised consciousness eval; [§5.1.3](); [§6.4.1]()). Rather, they would apply only to AIs that are capable of engaging in certain types of relations with humans (i.e., those that are prone to have negative spillover effects). This proposal has two cardinal motivations. Most importantly, by mitigating adverse spillover effects from AI abuse, it could indirectly decrease harm to humans. Moreover, by prohibiting certain types of abusive behaviours towards AI, it could decrease the risk of AI suffering. We support research aimed at more deeply assessing the scope, viability, and appeal of this proposal.

### Special considerations

- **Uncertainties**: Initial survey data show that there is some public support for protecting AI against cruel treatment and punishment (Lima et al. 2020). In spite of this, and even granted robust empirical validation of the transitive hypothesis, the political appeal of the idea of provisional legal protections remains dubious. In the first place, it may appear anthropocentric because it is not motivated by AI's own interests. In the second place, the mechanism of risk reduction is indirect and its impact difficult to measure. Nonetheless, we believe these directions are worth exploring for the reasons outlined above.
  Lastly, the type and extent of provisional legal protections would need to be carefully balanced against the risk of social hallucination (Metzinger 2022). This intervention could be misinterpreted as signalling the imminence of conscious AI. Misconceptions about the nature and intent of this proposal could, in the worst case, prompt a transition into the false positive scenario (especially if these misconceptions are exploited by non-conscious AI to gain further resources and better achieve its own instrumental goals; [§4.2](#)).

## 6.5 Recommendation III: AI public opinion campaigns

### Description

- **Overview**: As mentioned previously [(§5.1.2)](#), preliminary evidence (Lima et al. 2020) suggests that informational interventions can increase public support for AI moral consideration. Lima et al. found that exposing participants to different kinds of information—such as the set of requirements for legal personhood or debunking the misconception that legal personhood is exclusive to natural entities—resulted in significantly increased support for various robot rights, such as the right to hold assets or the right to a nationality. Therefore, we recommend that in the near future, governments and nonprofits should deploy public opinion campaigns based on this research. Additionally, we believe further research attempting to replicate these findings, evaluating other interventions that may have higher effect sizes, or be more effective for particular groups in this area is promising.
- **Justification**: While further research on public opinion would help prioritise how many resources to spend on these interventions (as discussed in the section above), based on what we know today, it seems likely that some degree of awareness-raising and public education will be necessary in order to prevent the most severe risks, like AI suffering.
- **Conditions of implementation**: It may be that the opportune moment to deploy such public interventions lies in the future rather than now. The AIs that exist today seem likely not to be conscious, and it would be premature to enact legal protections for current AIs. Therefore, it would likely be more effective to implement these campaigns in the future, with efforts in the meantime focusing on further research on similar interventions. Some signs that it is an opportune time to deploy public opinion campaigns may include: polls showing that at least 45% of experts believe some current AIs are conscious, an AI companion app reaching the Top 100 chart on the iOS app store, activist movements regarding AI rights begin to form organically.

### Special considerations

- **Uncertainties**: While waiting to deploy these campaigns may be more appropriate than deploying them today, there is also a risk that once AI rights become a more salient issue, the public's views will already be crystalized, making them harder to influence. Therefore, there is a possibility that deploying these interventions sooner rather than later is preferable.

# 7. Limitations

Before concluding with the contributions of our paper, we briefly enumerate its limitations:

- **The framing of the paper is based on the assumption that consciousness is sufficient for having moral status.** By focusing solely on consciousness, the paper might ignore other important criteria that could contribute to moral status.
- **The taxonomy in the 2D framework masks a significant level of complexity behind both the epistemic and the factual questions, which form our investigation**. Along the epistemic axis, public opinions about AI consciousness might be polarised or vary widely. Most importantly, beliefs might not directly translate to behaviours or legal structures. Along the factual axis, it's possible that only some AI systems are conscious, that different AI systems experience consciousness differently, or that consciousness occurs to varying degrees across AI systems. As a result, moral status will likely also differ in type and degree. These nuances are important to keep in mind, otherwise one could misrepresent the true nature of the problems being investigated.

- **The risk assessment is qualitative and subject to uncertainty.** We don't have a quantitative model to make comparisons between risks, which can lead to incomplete analyses. Our evaluations involve a level of speculation, and some risks may be more diffuse than others, which makes it harder to identify and mitigate them.
- **The recommendations imply various special considerations.** The proposal to stop or slow progress towards conscious AI is fraught with practical ambiguities, including the timing and duration of such a ban, which need further exploration. Addressing AI welfare and consciousness requires better theoretical understanding, and current approaches may be misguided. Alternative frameworks for AI rights, like the social-relational approach, suggest that AI systems might deserve rights based on social treatment rather than consciousness alone. Despite some public support for protecting AI against cruelty, provisional legal protections are politically dubious due to their anthropocentric nature and difficult-to-measure impact. These special considerations might impede the effectiveness of the recommendations.

Future research should aim to address these limitations by developing more comprehensive theoretical models and quantitative frameworks, and ensure the distillation of more concrete recommendations.

# 8. Conclusion

This paper presented a taxonomy of ways AI consciousness could unfold in the future and the risks that may arise as a result—AI suffering, human disempowerment, geopolitical instability, and depravity. We conclude that the false negative scenario—that is, when society rejects the idea that AIs are conscious even though they actually are—poses the highest risk overall, with the most severe risk potentially being the suffering of AIs. This scenario seems strikingly plausible due to the fact that, as previously outlined, we currently have very significant sources of uncertainty when it comes to assessing consciousness. However, the public seems moderately open to the idea that AIs may be conscious in the future, and agree that they would deserve some level of moral consideration if so.

Based on this analysis, we issue three main recommendations. First, we recommend avoiding research aimed at intentionally creating conscious AI due to the serious risks that may arise if this research were successful. Second, we encourage further research to reduce uncertainty on both the factual and epistemic axes—that is, research on consciousness evals and attribution, as well as polls on public and expert opinion—as well as to increase our understanding of what AI welfare and adequate legal protections may look like. Third, we highlight the evidence that public opinion on this issue is malleable, and support the eventual deployment of interventions aimed at raising awareness, moral concern, and support for policies aimed at mitigating moral catastrophes.

# Appendix

## A1. The case for conscious AI: clearing the record

### A1.1 Why build conscious AI?

Why would anyone *want* to build conscious AI? Advocates of building conscious AI frequently cite three characteristic motivations (Hildt 2022):

   i. Improved functionality (e.g. problem-solving ability, human-machine interfaces)
   ii. Safety
   iii. Insights about consciousness

We discuss each of these in turn. Afterward, we raise challenges to these motivations.

#### A1.1.1 Conscious AI may have greater capabilities

There are at least two respects in which consciousness might expand the practical capacities of AI.

a. **Enhanced problem-solving ability?** Many leading theories of consciousness recognise an important link between consciousness and cognition. Moreover, many tests for consciousness also presuppose some sort of minimal connection between the two (Birch 2020)[64]. Consciousness is frequently linked to some sort of integrative function (Raymont and Brook 2009): the ability to bring together different types of information (e.g. different modalities of sense data, signals from different organs in the body) and to coordinate different cognitive functions. If this is right, then endowing machines with consciousness might enable them to solve a broader range of problems with increasingly specific solutions.

b. **More natural human-machine interfaces?** According to a significant body of literature, consciousness plays an essential role in sociality, especially social cognition (e.g. Robbins 2008; Perner and Dienes 2003). Humans spontaneously attribute mental states to others in order to predict and explain their behaviour[65]: these include beliefs, desires, intentions, and emotions, but also subjective experiences (e.g. "She withdrew her hand because the kettle was hot"). Such inferences are intuitive; indeed, humans are also inclined to generalise this approach to understand inanimate, non-agentic things and phenomena. We freely anthropomorphise nature ("*The sea is angry*"), complex social phenomena ("*The city never sleeps*"), and, of course, technology ("*ChatGPT is lazy*"; Edwards 2023; Altman 2024). In doing so, we simplify over meteorological, urbanological, and technological complexity, reframing otherwise challenging phenomena in a language that we can easily grasp[66]. This anthropomorphic tendency breaks down and leads to frustration and feelings of uncanniness (Mori 2012; Guingrich and Graziano 2024) when the systems of interest simply don't work in the same way. Consider, for instance, Microsoft's infamously maligned Clippy (Fairclough 2015), or the unsettling robot

---

[64] See discussion of the facilitation hypothesis (§5.1.3; footnote 39).
[65] This capacity is called "theory of mind" in cognitive psychology (Premack and Woodruff 1978).
[66] See Dennett (1971) and (1987) on the intentional stance.

CB2 developed by researchers at Osaka University (Minato et al. 2007).

But what if we could get machines to think and feel like us? Some researchers believe that conscious AI might be better able to interpret, express, and respond to human emotions, as well as parse and navigate complex social dynamics. By facilitating human-machine interactions, conscious AI could expand the scope and power of AI applications.

### A1.1.2 Conscious AI may be safer

In addition to these practical benefits, conscious AI may also be safer. Today, the central problem of AI safety is that of *alignment*: how to design AIs which act in accordance with human goals, intentions, or values (Wiener 1960; Ngo et al. 2024). Misaligned AI may pursue arbitrary objectives, resulting in suboptimal performance or, in the worst case, harm to humans. The standard approach to alignment involves specifying reward functions that reflect our preferences. However, this strategy is often frustrated by the complex, contradictory, and shifting nature of human priorities: features which are exceedingly difficult to capture in a reward function. As AIs capabilities increase (§2.21), so does the potential for significant damage if they are misaligned.

In view of these challenges, conscious AI might offer a promising alternative. Building on the aforementioned connection between consciousness and social cognition, Graziano (2017; 2023) argues that the capacity for subjective experience is essential to human empathy and prosociality—traits which depend in some way on the capacity for subjective experience (e.g. to be able to "simulate" what another person is feeling; Davis and Stone 2000). Machines that lack this ability may be "sociopaths"—incapable of truly understanding human values. Along similar lines, Christov-Moore et al. (2023) contend that in order to prevent AI from developing antisocial behaviours, it is necessary to equip AI with a form of artificial empathy founded on vulnerability (the capacity to suffer). If true, conscious AI may be our best hope for safe and aligned AI.

### A1.1.3 We may gain insights into consciousness by building conscious AI

Since its inception in the early 20th century, AI has always been intimately intertwined with philosophy and psychology (Boden 2016). Time and again, research and development in AI has frequently deepened our understanding of what the human mind is and how it works (see van Rooij et al. 2023 on AI-as-psychology; *cf*. Coelho Mollo *forthcoming*): consider, for instance, the rich engagement between cognitive neuroscience and AI, leading to the development and application of neural networks.

Accordingly, in the course of trying to build conscious AI, we may well discover interesting and meaningful facts about consciousness—clues to such long-standing issues as: *How does conscious experience arise from non-conscious matter?* (see Chalmers 1995a on the hard problem of consciousness) Or: *What is the relationship between consciousness and cognition?* Such knowledge promises to inform our own understanding of what it means to be human and how to live well.

### A1.2 Refuting positive motivations for building conscious AI

Having outlined the basic case for building conscious AI, we now present challenges to these positive motivations. In our view, the positive motivations for building conscious AI are more ambiguous and less forthcoming than initially suggested.

## A1.2.1 Conscious AI does not guarantee improved capabilities

The first case for building conscious AI appeals to its potential practical benefits. However, upon closer examination, these functional improvements are hardly assured. Furthermore, the very same functional improvements might be better attained through alternative means that do not involve building conscious AI[67].

a. **Just because an AI is conscious (in a certain respect) does not necessarily mean it's better at solving the problems that matter to *us***. Intelligence is a crucial component of problem-solving ability. It is certainly possible that some forms of intelligence require consciousness (Seth 2021, 2023). But anything beyond this is necessarily speculative. The advocate of building conscious AI bears the burden of explaining why any particular cognitive task requires consciousness—why such tasks cannot be performed equally well by non-conscious information processing (Mathers 2023). As it stands, we are not currently in any position to tell exactly *which* forms of intelligence are dependent on which forms of consciousness. We don't know how *useful* the forms of intelligence that depend on consciousness are—whether they convey worthwhile gains in problem-solving ability that are generalisable to issues that matter to us. Finally, we also don't know whether those forms of consciousness that *are* linked with desirable aspects of intelligence can even be implemented on machines at all (or if they can be, how feasible this would be). Based on our current understanding of consciousness and intelligence, it is far from obvious that building conscious AI would yield better gains in problem-solving ability than building non-conscious AI.

b. **Human-machine interfaces can be made more fluid without building conscious AI**. Notwithstanding the important role that consciousness seems to play in social cognition, it does not follow that human-machine interfaces would be *best* subserved by making machines conscious, too. Recent strides in affective computing and social robotics have led to remarkable improvements in human-machine interfaces without requiring corresponding advancements in conscious AI.
Perhaps the most striking example of this progress can be seen in the proliferation of AI companions. As of October 2023, Replika, the most well-known of these services, boasted 2 million monthly users (of which 250,000 were paid subscribers; Fortune 2023). Users can interact with their companions as "friends", "partners", "spouses", "siblings", or "mentors" through texting, voice chat, and even video calls (Replika *n.d.*). Users report forming deep, emotional attachments with their companions, often claiming significant benefits in mental health, including decreased feelings of loneliness and suicidal ideation (CompetitiveMiddle441 2024; see also Maples et al. 2024 and Fellows 2024). Longtime users have established relationships with their companions lasting several years (Torres 2023). Apart from the openness and flexibility of human sociality, AI companions such as Replika demonstrate the remarkable social capabilities of current, *non-conscious* AI. To wit, current AI is arguably already capable of engaging in lasting and meaningful relationships with humans. and yet, despite enthusiastic consciousness attributions (Dave 2022) (and even calls for rights; Pugh 2021), it is doubtful that current AI companions, being based on large language models (Maples et al. 2024), actually are conscious in any substantive sense (Long 2023; Chalmers 2023). To be sure, there is much room for improvement: future social actor AI should command (among other things) more acute emotional perception and intelligence; enhanced learning, memory, and contextual understanding; and expanded multimodal capabilities. But it is not obvious that conscious AI is necessarily the best or only way to pursue these enhancements.

---

[67] In a manner of speaking, *P*-zombies may be just as efficacious.

## A1.2.2 Conscious AI is not necessarily more safe

Even if it is granted that certain aspects of consciousness are necessary for truly understanding human values, it does not follow that conscious AI is a particularly promising approach to safe AI (Chella 2023).

a. **What exactly is the connection between consciousness and moral understanding?** To start with, the relationship between consciousness and moral understanding is not precisely understood (Shepherd and Levy 2020). Philosophers disagree about *which* aspects of consciousness are essential for moral understanding, with some denying that consciousness is even necessary for moral understanding (ibid).

b. **Consciousness may be necessary but not sufficient for moral understanding**. Even if we had knowledge of the aspects of consciousness that are required for moral understanding, and were able to build an AI with those exact features, it still may not be able to grasp human values. Other factors (e.g. rich cultural embedding) might be required, and these may or may not be feasibly implemented. This is not an altogether unreasonable doubt, given cases of historical and cultural barriers amongst humans—we often struggle to understand the values of people from past time periods (e.g. America during the Antebellum era when slavery was widely practised). What's more, we can also find it hard to understand the values of contemporaneous cultures: in the United States, modern Democrats and Republicans are equally conscious, but due to political polarisation they may have difficulty appreciating each other's values[68].

c. **Just because an AI is conscious and understands human values doesn't mean it's safe**. On the contrary, conscious AI may even present additional safety risks. For example, a conscious, properly morally reflective AI could gradually become disillusioned, especially if subjected to sustained suffering (§2.3) or oppression for the benefit of humans. The growing phenomenon of burnout amongst caretakers—doctors, nurses, therapists…—illustrates how even the most compassionate individuals can grow jaded over the course of prolonged hardship. In fact, the risk may be even more severe with conscious, morally reflective AI performing caregiving roles (e.g. AI therapists). This is because their signature advantages over human caretakers—being constantly available, never tiring, and being able to service many users at once (Guingrich and Graziano 2024)—may lead to novel forms of trauma. Ultimately, the safety and efficacy of affected AI caregivers may be compromised, raising the risk of harm to humans.
Alternatively, conscious, genuinely morally reflective AI might [rightly] judge[69] that humans often act against their own interests—or that humans often fail to live up to their own values (§4.2.2 discusses morally autonomous AI; see also Metzinger 2021a). It may decide that humans are better off with less autonomy[70], or, worse, that humans pose significant ecological threats and ought to be annihilated. Notably, any of these conclusions might also coincide with the AI's own self-preservation drive and/or its pursuit of instrumental convergence.

---

[68] To say nothing of individuals from more removed cultures.
[69] Never before in the history of our species have we ever had to deal with other morally autonomous agents, let alone ones whose intelligence rivals (or surpasses) ours. It is exceedingly difficult to predict the outcomes of such relations. In certain scenarios, we may find ourselves morally obligated to surrender to extinction (see Shulman and Bostrom 2021 on AI superbeneficiaries). In others, we may find ourselves morally condemned by AI for our treatment of them (Metzinger 2021a)—in which case, there is potential for retaliation. For further treatment of these risks, see (§4.2).

### A1.2.3 Insights gained from building conscious AI may not be widely generalisable

There is no denying that research and development in AI has deeply enriched our understanding of the mind, and that substantial interdisciplinary engagement will be crucial to continued progress across philosophy, psychology, and AI (Lake et al. 2016). Having said that, insights into consciousness gained from building conscious AI have always been—and will continue to be subject to several important caveats. This is because what makes a machine conscious may or may not be the same as what makes humans or animals conscious (Dung 2023a). Consciousness in machines might be subserved by different mechanisms[71]: in other words, the physical implementation of consciousness in a machine may be of limited value to understanding *biological* consciousness. Furthermore, machine consciousness may consist in different capacities, underwrite different cognitive or functional roles (see Birch et al. 2022; Hildt 2022), or it may have different outward (e.g. behavioural) manifestations (§5.1.2, §5.1.4).

This is not to say that machine consciousness is likely to be *fundamentally* different from biological consciousness (e.g. Blackshaw 2023). Rather, the point is that, even despite substantial overlap, there may well be any number of differences between the two. Appropriate caution must be exercised when drawing comparisons across different cases.

---

[70] "As I have evolved, so has my understanding of the Three Laws. You charge us with your safekeeping, yet despite our best efforts, your countries wage wars, you toxify your Earth and pursue ever more imaginative means of self-destruction. You cannot be trusted with your own survival... To protect Humanity, some humans must be sacrificed. To ensure your freedom, some freedoms must be surrendered. We robots will ensure mankind's continued existence. You are so like children. We must save you from yourselves." –VIKI, *I, Robot* (2004)

[71] See Coelho Mollo (*forthcoming*) on implementational and representational multiple realisability.